\documentclass[10pt,journal,compsoc]{IEEEtran}

  \usepackage{cite}

\IEEEoverridecommandlockouts

\usepackage{caption}
\usepackage{amsmath,amssymb,amsfonts}
\usepackage{algorithmic}
\usepackage{graphicx}
\usepackage{textcomp}
\usepackage{xcolor}
\usepackage{listings}
\usepackage{url}
\usepackage{cite}
\usepackage{color}
\usepackage{commath}
\usepackage{multirow}
\usepackage{balance}
\usepackage{cite}
\usepackage{comment}
\usepackage[shortlabels]{enumitem}

\newcommand{\inlinedComment}[2]
{\textcolor{#1}{\small\textbf{#2}}}

\newcommand{\rev}[1]{\textcolor{black}{#1}}
\newcommand{\minor}[1]{\textcolor{black}{#1}}
\begin{document}

\title{\huge Checking Smart Contracts with \\ Structural   Code Embedding}
\author{Zhipeng Gao, Lingxiao Jiang, Xin Xia, David Lo and John Grundy
 \IEEEcompsocitemizethanks{
 \IEEEcompsocthanksitem Zhipeng Gao, Xin Xia and John Grundy are with the Faculty of Information Technology, Monash University, Melbourne, Australia. \protect\\
 E-mail: \{zhipeng.gao, xin.xia, john.grundy\}@monash.edu
 \IEEEcompsocthanksitem Lingxiao Jiang, David Lo are with the School of Information Systems, Singapore Management University, Singapore.\protect\\
 E-mail: \{lxjiang, davidlo\}@smu.edu.sg 
\IEEEcompsocthanksitem Xin Xia is the corresponding author.}
 \thanks{Manuscript received     ; revised   }}

%

\markboth{IEEE Transactions on Software Engineering, ~Vol.~  , No.~  , }%
{Shell \MakeLowercase{\textit{et al.}}: Bare Demo of IEEEtran.cls for Computer Society Journals}

\IEEEtitleabstractindextext{%
\begin{abstract}
Smart contracts have been increasingly used together with blockchains to automate financial and business transactions. However, many bugs and vulnerabilities have been identified in many contracts which raises serious concerns about smart contract security, not to mention that the blockchain systems on which the smart contracts are built can be buggy. Thus, there is a significant need to better maintain smart contract code and ensure its high reliability.
\rev{
In this paper, we propose an automated approach to learn characteristics of smart contracts in Solidity, which is  
useful for clone detection, bug detection and contract validation on smart contracts.} 
Our new approach is based on word embeddings and vector space comparison. We parse smart contract code into word streams with code structural information, convert code elements (e.g., statements, functions) into numerical vectors that are supposed to encode the code syntax and semantics, and compare the similarities among the vectors encoding code and known bugs, to identify potential issues.
We have implemented the approach in a prototype, named {\sc SmartEmbed}
\footnote{The anonymous replication packages can be accessed at: \url{https://drive.google.com/file/d/1kauLT3y2IiHPkUlVx4FSTda-dVAyL4za/view?usp=sharing}}, and evaluated it with more than 22,000 smart contracts collected from the Ethereum blockchain. Results show that our tool can effectively identify many repetitive instances of Solidity code, where the clone ratio is around 90\%. Code clones such as type-III or even type-IV semantic clones can also be detected accurately.
Our tool can identify more than 1000 clone related bugs based on our bug databases efficiently and accurately.
Our tool can also help to efficiently validate any given smart contract against a known set of bugs, which can help to improve the users' confidence in the reliability of the contract.

\end{abstract}

}

\maketitle
\IEEEdisplaynontitleabstractindextext

\section{Introduction}
\label{sec:intro}
A \emph{Smart Contract}, a term coined by Nick Szabo in 1994 \cite{Szabo1994}, is a program that can be triggered to execute any task when specifically predefined conditions are satisfied.
The conditions defined in smart contracts, and the execution of the contracts, are supposed to be trackable and irreversible in such a way that minimizes the need for trusted intermediaries. They are also supposed to minimize either malicious or accidental exceptions in order to ensure trustworthiness of any business transactions implied by the smart contracts.

In recent years, along with widely-deployed cryptocurrencies (e.g., Bitcoin, Ethereum, and many others) on distributed ledgers (a.k.a., blockchains), smart contracts have  obtained much attention and have been applied to many business domains to enable more efficient and trustable transactions. The overall market capitalization of cryptocurrencies is more than 200 billions in USD as of August 2018 \cite{CoinMarketCap2018}. Many crytocurrencies involve various kinds of smart contracts, and a smart contract in the blockchains often involves cryptocurrencies worthy of millions of USD (e.g., DAO \cite{DAO2018}, Parity \cite{Parity2017} and many more). This gives much incentive to hackers for discovering and exploiting potential problems in smart contracts, and there is a very significant need to check and ensure the robustness of smart contracts.

Even though there have been many studies on the characteristics of bugs in smart contracts and underlying blockchain systems (e.g., \cite{Wan2017,li2017survey,atzei2017survey,bartoletti2017empirical,chen2017under}) and detection of smart contract bugs (e.g., \cite{Bhargavan2016,brownformal,luu2016making,tsankov2018securify,tikhomirov2018smartcheck,delmolino2016step,Mueller2018}),
there are still increasing needs to detect and prevent more and more kinds of problems identified in smart contracts.
A major disadvantage of these existing bug detection tools is that they require certain bug patterns or specification rules defined by human experts in order to construct bug detectors and/or code model checkers to check smart contracts against the defined rules. With the high stakes in smart contracts and race between attackers and defenders, it can be far too slow and costly to write new rules and construct new checkers in response to new bugs and exploits created by attackers.

In this paper, we propose a new approach that addresses the above issue. We aim to enable efficient checking of smart contracts and can evolve checking rules along with the evolution of code and/or bugs, based on our deep learning model for smart contracts.
The main idea of our approach is two fold:
(1) code and bug patterns, including their lexical, syntactical, and even some semantic information, can be automatically encoded into numerical vectors via techniques adapted from word embeddings (e.g., \cite{ye2016word,bojanowski2016enriching,mikolov2013efficient,mikolov2013distributed,turian2010word}) enhanced with basic program analyses and the availability of many smart contracts;
(2) code checking can be essentially done through similarity checking among the numerical vectors representing various kinds of code elements of various levels of granularity in smart contracts. This idea, with suitable concrete code embedding and similarity checking techniques, can be general enough to be applied for various code debugging and maintenance tasks. These include repetitive (a.k.a.~duplicate or cloned) contract detection, detection of specific kinds of bugs in {\em a large contract corpus}, or validation of a contract {\em against a set of known bugs}\footnote{\rev{``Validation''  in this paper is to check if a contract has {\em no bug similar to the known bugs}; it does not mean formal verification of the contract.}}.

We have built a prototype based on the idea, named {\sc SmartEmbed}, for smart contracts written in the Solidity programming language \cite{Solidity2018} used in the Ethereum blockchain \cite{EtherScan2018}. We have collected 22,725 contracts in their Solidity source code that are labelled as ``verified'' in the Ethereum blockchain and 17 well-known buggy contracts from the Internet. Our tool can then automatically generate the vector embeddings from the contract code collected from the blockchain and provides a mechanism to compose vector embeddings for any code fragment, either buggy or correct. All of these vectors then go through similarity checking for different purposes. Our evaluation results against 22,725 contracts show that, for the tasks of clone detection, bug detection, and contract validation,  our approach can achieve comparable results compared with specific tools such as Deckard\cite{jiang2007deckard}, SmartCheck\cite{tikhomirov2018smartcheck}.


The main contributions of this paper are as follows:

\begin{itemize}
	\item We propose a new approach for Solidity code checking based on code embedding and similarity checking, which is applicable for various purposes, such as similar contract code detection, bug detection, and contract validation.
	
	\item We built a prototype {\sc SmartEmbed} based on the approach, and evaluated it on more than 22,000  Solidity contracts collected from the Ethereum blockchain. 
	
	\item Our clone detection results show that our tool can effectively identify many repetitive Solidity code where the clone ratio is around 90\%, and we can detect \rev{more semantic clones accurately} than the commonly used clone detection tool Deckard.
	
	\item
	\rev{
	 Our bug detection results show that {\sc SmartEmbed} can identify more than 1,000 clone related bugs based on our bug databases efficiently and accurately, which can enable efficient checking of smart contracts with changing code and bug patterns. For contract validation, our approach can capture bugs similar to known ones with low false positive rates, the query for a clone or a bug is quite efficient which can be sufficient for practical uses.
	}
\end{itemize}

This paper is organized as follows. Section~\ref{sec:related} presents related work on smart contract security and relevant techniques. Section~\ref{sec:approach} presents our approach for smart contract code embedding. Section~\ref{sec:eval} evaluates our approach on actual contracts collected from the Ethereum blockchain. Section~\ref{sec:threats} discusses limitations of our approach and its evaluation. Section~\ref{sec:con} concludes the paper.

\section{Related Work}
\label{sec:related}

\begin{figure*}
\centerline{\includegraphics[width=1\textwidth]{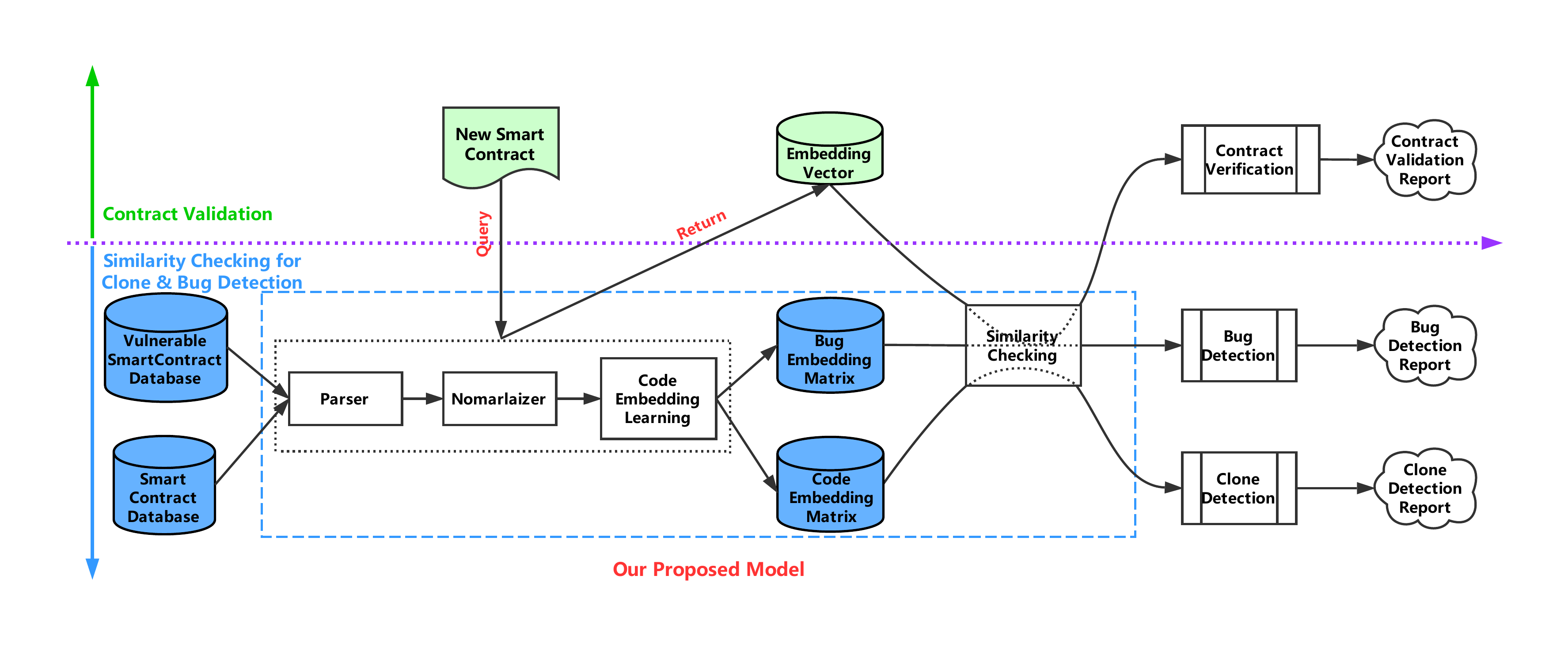}}
 \vspace*{-35pt}
\caption{Overview of Our Approach}
\label{fig:approach}
\end{figure*}


\subsection{Smart Contract and Security Problems}
Despite the fact that Ethereum and smart contracts are relatively new, many studies have been performed on security aspects of smart contracts.
Some studies focus on creating taxonomies of smart contract security vulnerabilities (e.g., \cite{alharby2017blockchain,egbertsen2016replacing,zheng2016blockchain,delmolino2016step}).
Others focus on specific bug detection. For example,
Loi et al.~\cite{luu2016making} build a symbolic execution tool called OYENTE to detect four kinds of security bugs.
Tikhomirov et al.~\cite{tikhomirov2018smartcheck} build a static analysis tool called SmartCheck to automatically check for vulnerabilities and
code smells.
Brown et al.~\cite{brownformal} present a framework for analyzing runtime safety and functional correctness of smart contracts via formal verification; several types of vulnerability, such as reentrancy and exception disorders, can be identified by their tool.
Chen et al.~\cite{chen2017under} developed a security tool for identifying gas costly programming patterns in smart contracts.

Although the aforementioned research has proposed security analysis tools to find bugs in smart contracts, most of those tools are built to discover specific types of potential vulnerabilities, requiring manually constructed bug patterns or specifications. To the best of our knowledge, no one has yet considered how to make such tools more flexible and adaptive to arbitrary new bugs by using word embedding for smart contract code. Our work is the first to propose an approach for detecting smart contract bugs and validating contracts via similarity checking of contract code embeddings, \rev{especially the embeddings that take code structures into consideration}.


\subsection{Word Embedding and Code Similarity}
Embedding (also known as distributed representation \cite{mikolov2013distributed, turian2010word}) is a technique for learning vector representations of entities such as words, sentences and images. One of the typical embedding technique is word embedding, which represents each word as a fixed-size vector, so that similar words are close to one another in the vector space~\cite{mikolov2013efficient,mikolov2013distributed,ye2016word,bojanowski2016enriching}.

Recently, an interesting direction in software engineering is to use deep learning to compute and use vector representations of programs.
For example, Mou et al.~\cite{mou2014building} propose to learn vector representations of source code. They map the nodes of abstract syntax trees to vectors.
Following their previous work, Mou et al.~\cite{mou2016convolutional} propose a tree-based convolutional neural network based on program abstract syntax trees to detect similar source code snippets.
Ye et al.~\cite{ye2016word} embed words into vector representations to score a pair of documents, and use StackOverflow questions and answers as document corpora to train word embeddings.
White et al.~\cite{white2017sorting} propose an automatic program repair approach, DeepRepair, which leverages a deep learning model to identify similarity between code snippets.

Different from these existing tools, our code embedding methods are based on serialization of solidity parse tree for different level program elements. To the best of our knowledge,
our work is the first to apply the code embeddings to the specific domain of Ethereum smart contracts as inspired by the promising results of employing deep learning to the many other software engineering tasks (e.g., \cite{white2015toward, corley2015exploring, yang2015deep, lam2015combining, white2016deep, wang2016automatically, gu2016deep}).

\subsection{Clone Detection, Bug Detection, and Code Validation}
A plethora of approaches have been investigated for different tasks such as code clone detection, bug detection, and code validation and/or program verification. All of the tasks can be viewed as variants of the problem of finding ``similar'' code, depending on the definition of similarity: code clone detection is to search for code in a code base ``similar'' to a given piece of code; bug detection is to search for code in a code base ``similar'' to a known bug; and code validation is to search for (non-existence of) code in a code base ``similar'' to any bug.
As our approach based on code embedding and similarity checking is an instantiation of this general view, it is related to many such studies too.

For clone detection, many techniques in the literature generally begin by generating some intermediate representations for code before measuring similarity. According to source code representation, these techniques can be classified as text-based (e.g., \cite{johnson1993identifying, johnson1994visualizing, ducasse1999language}),
token-based (e.g., \cite{baker1993program, baker1996parameterized, inoue2002ccfinder}),
tree-based (e.g., \cite{jiang2007deckard, koschke2006clone, yang1991identifying}),
graph-based (e.g., \cite{gabel2008scalable, komondoor2001using, krinke2001identifying, liu2006gplag}),
semantic-based (e.g., \cite{Jiang2009,Kamiya2013,Su2016,Kim2018}), deep-learning-based (e.g., \cite{white2016deep,Gu2018}), or a mixture.
Our approach complements those studies by applying word embedding to smart contract code and its syntax structures to search for smart contracts of various levels of granularity.

For bug detection, there also exists many conventional techniques tailored for smart contracts, such as those based on
static analysis and model checking (e.g., SmartCheck \cite{tikhomirov2018smartcheck}, Securify \cite{tsankov2018securify}),
symbolic execution and dynamic analysis
(e.g., Oyente \cite{luu2016making}), Manticore \cite{TrailOfBits2017}), and a mix of techniques (e.g., Mythril \cite{Mueller2018}). ``Conventional'' here refers to the fact that they require human curated correctness and/or bug patterns or specifications in order to check whether the code complies with or violates the given patterns or specifications.

There are other bug detection techniques that do not require predefined bug patterns or specifications; instead, they often rely on statistically inconsistencies among multiple instances of code. For example,
Juergens et al.~\cite{haque2018causes} report that inconsistencies among similar code are an important source of bugs in programs, and every second (possibly inconsistent) modification of a piece of similar code increases the chance of errors.
This phenomenon has been explored in the literature to detect clone-related bugs (e.g., \cite{jiang2007context,li2004cp}), code porting errors (e.g., \cite{Ray2013}), semantic bugs (e.g., \cite{Dillig2007,lawall2010finding,Srivastava2011}), etc.

Another category of bug detection techniques depending on historical known bugs is more similar to our approach. Those approaches learn patterns from known bugs using various techniques (e.g., graph pattern matching \cite{Li2012} and heuristic rule matching \cite{jiang2007context,li2004cp}) and search for similar instances in a given code base.
Recently, such techniques that require little or zero efforts in manually written specifications are often based on deep learning (e.g., \cite{Yang2015,Pradel2017}).

Our approach is relying on the existence of known bugs,
as it automatically learns code and bug representations from known bugs based on code embedding.
It is unsupervised; there is no need to handcraft features beforehand, which saves much manual effort in feature selection needed for many other techniques.
Given a sufficiently comprehensive set of code and known bugs, our approach can potentially be applicable for both bug detection and contract code validation.
\rev{On the downside, our ``bug detection'' and ``contract validation'' are both evaluated {\em with respect to the known bugs}: bug detection is to detect all instances of the known kinds of bugs in a large contract corpus; contract validation is to check if a contract is free of any instance of bugs similar to the known bugs. 
If no enough known bugs are available, our approach can utilize potential bugs reported by conventional techniques too, providing a complementary way to make bug detection and contract validation more comprehensive.
}


\section{Approach}
\label{sec:approach}

\begin{figure*}
\centerline{\includegraphics[width=0.95\textwidth]{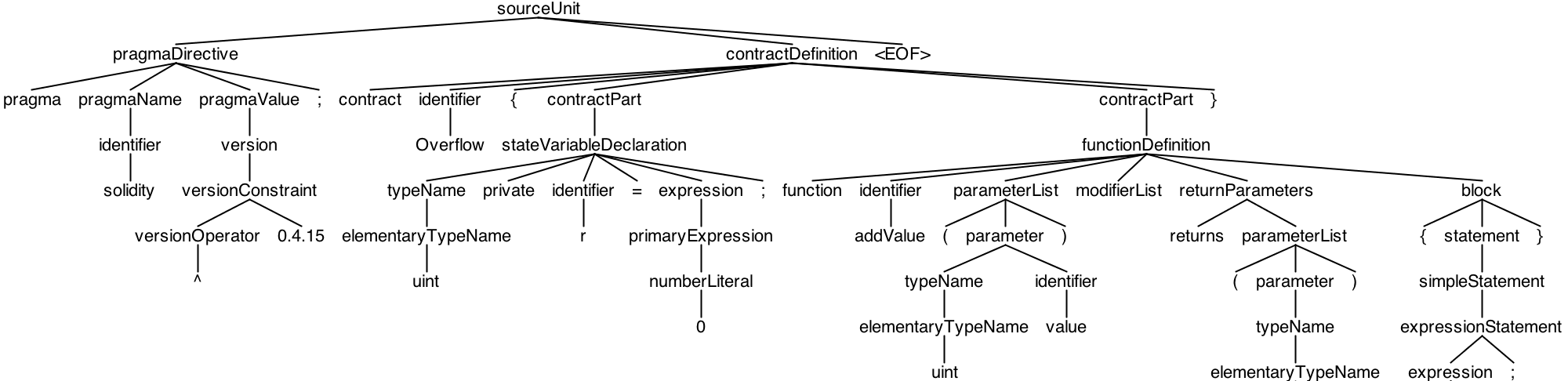}}
\vspace*{-2pt}
\caption{Sample Solidity Parse Tree}
\label{fig:parsetree}
\end{figure*}


Fig.\ref{fig:approach} demonstrates the overall framework of {\sc SmartEmbed}. Based on similarity checking and code embeddings, {\sc SmartEmbed} is targeting three tasks: clone detection, bug detection, and contract validation.
For clone detection and bug detection, we aim to identify code clones and clone-related bugs for smart contracts in the existing Ethereum blockchain. For contract validation, given a new smart contract, {\sc SmartEmbed} will help to validate whether it contains vulnerable statements associated with our bug database.

To be more specific,
the collected source code of smart contracts are loaded and parsed by our custom built parser, generating the abstract syntax trees (ASTs) for a smart contract. Then, we extracted a stream of tokens by serializing the ASTs. Following that, the normalizer reassembles the token stream to eliminate the differences (e.g., the stop word, values of constants or literals) between smart contracts. The result sequence that is output by the normalizer is then fed into our code representation learning sub-model. Through the model building and training, each code fragment would be embedded by a fixed-length dimension vector. All of the source code will be encoded into the code embedding matrix. In the meanwhile, all vulnerable source code would be embedded into the bug embedding matrix.

Next, clone detection, bug detection and contract validation are performed using similarity checking methods via vector space comparison.
Similarity comparison is performed between the possible code snippet pairs, and a similarity threshold governs whether code fragments will be considered as code clones or clone-related bugs.

In following sub-sections, we elaborate our data collection, parsing, normalization, embedding learning, and similarity checking steps.

\subsection{Data Collection}\label{AA}

\rev{To prepare the smart contract code used for our approach and evaluation,}
firstly we collected Solidity smart contracts using EtherScan\footnote{https://etherscan.io/}, which is a block explorer and analytics platform for Ethereum. To be more specific, we built our own web scrapers to systematically search and download every HTML page on the entire site. After parsing HTML output from that page,  needed information (e.g. contract address/source code/byte code/opcodes) were extracted from the HTML file for our further assessment.

By April 20, 2018 when we started our evaluation experiments, we had collected 22,725 verified smart contract. We counted the number of individual contracts (given the source code of a smart contract, there may include several individual contracts), functions, statements, and lines associated with these smart contracts. On average, each smart contract involves around 6 individual contracts, 27 functions, 85 statements, and 323 lines of code. Table~\ref{tab:dataoverview} describes the statistics of our collected dataset.

\begin{table} 
\caption{Collected Data}
\begin{center}
\minor{
 \begin{tabular}{|c|r|}
    \hline
    \# Contracts & 22,725 \\
    \hline
    \# Individual Contracts & 135,239 \\
    \hline
    \# Functions & 631,261   \\
    \hline
    \# Statements & 1,944,513 \\
    \hline
    \# Lines of Code & 7,329,362 \\
    \hline
\end{tabular}
\label{tab:dataoverview}
}
\end{center}
\end{table}

\subsection{Parsing}
The abstract syntax tree (AST) is a structural representation of a program. In this step, for each smart contract, we used a custom-built Solidity parser to parse the smart contract into an AST. We built our code embeddings based on AST because its tree structured nature provides opportunities to capture structural information of programs.

More specifically, ANTLR and a custom Solidity grammar were used to generate the XML parse tree as an intermediate code representation. The source code was fully translated to this internal tree representation. After that, we built the code embeddings based on this abstract syntax tree. Listing 1 and Fig. \ref{fig:parsetree} provides a simple example of a smart contract and its corresponding AST, defined in Solidity.

We serialized the parse tree of a smart contract differently for contract-level, function-level and statement-level program elements, depending on the types of the tree nodes that contain or are siblings of the relevant elements. The high level idea of such a processing is to capture the structural information (e.g., branch and loop conditions) in and around the focal elements.
Further, non-trivial tokens and identifier names are processed and put into the code element sequences serialized from the trees, so that certain data flow information (via defining/using a same name) is added into the sequences too.
We describe the details of the tokenization process below with the aforementioned sample Solidity code.


\definecolor{codegreen}{rgb}{0,0.6,0}
\definecolor{codegray}{rgb}{0.5,0.5,0.5}
\definecolor{codepurple}{rgb}{0.58,0,0.82}
\definecolor{backcolour}{rgb}{0.96,0.96,0.96}

\lstdefinestyle{mystyle}{
  backgroundcolor=\color{backcolour},   commentstyle=\color{codegreen},
  keywordstyle=\color{magenta},
  numberstyle=\tiny\color{codegray},
  stringstyle=\color{codepurple},
  basicstyle=\footnotesize,
  breakatwhitespace=false,
  breaklines=true,
  captionpos=b,
  keepspaces=true,
  numbers=left,
  numbersep=5pt,
  showspaces=false,
  showstringspaces=false,
  showtabs=false,
  tabsize=2
}
\lstset{style=mystyle}
\title{Code Listing}
\begin{lstlisting}[language=Java, caption=An Example of Solidity Program]
pragma solidity ^0.4.15;

contract Overflow {
    uint private r=0;

    function addValue(uint value) returns (bool){
        // possible overflow
        r += value;
    }
}
\end{lstlisting}



\vspace{0.1cm}\noindent {\em  \bf Contract Level Tokenization:} We extracted all terminal tokens from the XML parse tree by performing an in-order traversal. Regarding the previous smart contract, the following tokens were extracted (1\_10 stands for the line range of this contract).
\begin{lstlisting}
1_10 : pragma solidity ^ versionliteral ; contract Overflow { uint private r = 0 ; function addValue ( uint value ) returns ( bool ) {  r += value ; } }
\end{lstlisting}

\vspace{0.1cm}\noindent {\em \bf Function Level Tokenization:} Considering the function level tokenization, we appended the contract signature to the end of function tokens.
For the previous smart contract, function level tokenization’s result was given as follows (6\_9 represents this function starts at line 6 and ends at line 9).
\begin{lstlisting}
 6_9 :  function addValue ( uint value ) returns ( bool ) { r += value ; } contract Overflow overflow { }
\end{lstlisting}

\vspace{0.1cm}\noindent {\em \bf Statement Level Tokenization:}
Different from the contract-level and function-level tokenization, for statement-level tokenization, based on the terminal tokens, we added more details of structural and semantic relations.
For example, regarding the previous smart contract, structural information such as the chain of ancestors in ASTs
as well as function signatures were retrieved from the XML parse tree.
\minor{By adding the chain of ancestors in ASTs, our model can capture the structural relationship; by adding the diverse neighbourhood nodes, our model can capture the ``context'' information of a focal element.
}
\begin{lstlisting}
 8_8 : sourceUnit contractDefinition contractPart functionDefinition block statement simpleStatement r += value ; function addValue add value ( uint value ) returns ( bool ) contract Overflow overflow { }
\end{lstlisting}

Our parse tree based serialization of the code with respect to a focal element captures most structural (containment and neighbouring) and some semantic (data-flow) information, which serves the downstream applications.

\subsection{Normalization}
An important task during preprocessing is normalization. In this step, we normalized the token sequence to remove some semantic-irrelevant information. To be more specific, the following steps have been taken:
\begin{itemize}
    \item Stop words : \rev{For single-character variables}, such as ``i'', ``j'', ``a'', ``b'', ``k'', etc., we replaced them with ``SimpleVar''. The below code snippet illustrates this step :
    \begin{lstlisting}
    uint private r = 0 ;
    ==>
    uint private SimpeVar = 0 ; \end{lstlisting}
    \item Punctuations : Tokens having no effect on code operational semantics, non-essential punctuations such as \textquoteleft, \textquoteright, ``,'', ``;'' were removed. Some other punctuations were reserved such as ``\{'', ``\}'', ``['', ``]''. The following code snippet exemplified this operation :
    \begin{lstlisting}
     uint private SimpeVar = 0 ;
     ==>
     uint private SimpeVar = 0 \end{lstlisting}
    \item Constants : According to the type of constants, we unified them with ``StringLiteral'', ``DecimalNumber'', ``HexNumber'' and ``HexLiteral'' respectively. The below gives an example of how this step works :
    \begin{lstlisting}
     uint private SimpeVar = 0
     ==>
     uint private SimpeVar = decimalnumber \end{lstlisting}
     \item Camel Case Identifiers : For  identifiers following camel casing, we kept it as a reserved token. Additionally, we split this identifier into its constituent individual words. For example,
     \begin{lstlisting}
      addValue
      ==>
      addValue add value \end{lstlisting}
\end{itemize}
The normalizer generated token stream of the 22,725 contracts, 631,261 functions and 1,944,513 statements respectively.
After the normalization process, 1.2GB of clean text remained, amounting to 119,568  tokens.
This comprised the final training dataset that was fed into the training algorithm.

\subsection{Code Embedding Learning}
In this step, based on the previous normalization results, we mapped each possible code fragment, such as statement, function, and contract to a high dimensional vector respectively. The following two embedding algorithms are applied: Word2Vec\cite{mikolov2013efficient} and FastText\cite{bojanowski2016enriching}. Word2Vec learns vector representations of words that are useful for predicting the surrounding words in a sentence. However, traditional Word2Vec failed to capture the morphological structure of a word. FastText attempts to solve this by treating each word as the aggregation of its subwords, subwords are taken to be the n-gram of the word, and the vector for a word with FastText is the sum of all n-gram vectors of its component.

To train the model, we used the open source Python library gensim\footnote{https://radimrehurek.com/gensim/}, which incorporates the Word2Vec and FastText training algorithm at the same time. We have to clarify that we choose FastText as our primary embedding methods for the later experiment because of the following reasons: 1) According to our experimental result, FastText performs better on syntactic tasks compared to the original Word2Vec. The reason for this may be that FastText take into account subword information, which captures more semantic and syntactic information from the context 2) FastText can be used to obtain vectors for out-of-vocabulary (OOV) words, by summing up vectors for its component char-ngrams. \rev{Since the number of unique tokens in the training dataset was very limited}, i.e. 119,568, OOV problems could be encountered very often when dealing with a new smart contract. The details of the code embedding learning process are described as follows:

\subsubsection{\textbf{Token Embedding}}
The normalized token stream generated by the normalizer was used as the training corpus. We then applied the embedding algorithm to contract-level, function-level, and statement-level training corpus respectively. After that, each token within the training corpus was mapped to a real-valued vector of a fixed dimension.
\rev{
Since there are 308 node types in Solidity’s grammar file,  we set the word embedding size to half of the number of node types, which is 150, for compressing irrelevant or overlapping meanings of the node types when SmartEmbed generates embeddings for the code.\footnote{\rev{Dimensions in the range of a few hundreds have been used in the literature \cite{ye2016word, gaoautomating, mikolov2013efficient, bojanowski2016enriching} with reasonably good effectiveness. We leave the sensitivity analysis of vector dimensions as future work.}}
}
The token embeddings process served as a “pretraining” stage for constructing higher-level code embeddings.

\subsubsection{\textbf{Higher Level Embedding}}
As long as we got basic vector representation for tokens, the embeddings of higher level code fragments such as statement-level, function-level, and contract-level were able to be generated by the composition of the possible atomic tokens.
To capture the features of semantics as well as the size of the code, we chose the summing metric to compose this shared embeddings in our preliminary study. Specifically, the code embeddings for a particular code fragment is summing up all possible tokens' embeddings within it. The more formal definition for the code embedding is described as follows:

\vspace{0.1cm}\noindent\textbf{Definition} : Given a solidity code snippet $T$, for each token $w$ in $T$, we define the code embedding for $T$ as following:
\begin{equation}
Embedding(T)= \sum_{w \in T} w_{vector} \label{eq1}
\end{equation}
After defining the code embedding for a particular code fragment, every possible smart contract, function, and statement can be embedded to a fixed-length vector.

\subsection{Embedding Matrix Building}
By stacking every single vector together, we can easily obtain 3 code embedding matrices $\mathbf{C^{c \times d}}$, $\mathbf{F^{f \times d}}$, $\mathbf{S^{s \times d}}$ with respect to contract-level, function-level, and statement-level respectively.

\vspace{0.1cm}\noindent\textbf{Contract Embedding Matrix} $\mathbf{C^{c \times d}}$: For contract-level code embedding matrix, the first dimension $c$ is the total number of contracts, which was 22,725, the second dimension $d$ is the code embedding size we set previously, which was 150 in our case. In other words, contract embedding matrix $\mathbf{C}$ would be a 22,725 $\times$ 150 matrix. We considered the $i$th element  $\mathbf{C_i}$ ($i = 1, 2, ..., c$), which is a 150 dimensional vector, as the code embedding for $i$th contract.

\vspace{0.1cm}\noindent\textbf{Function Embedding Matrix} $\mathbf{F^{f \times d}}$: For function-level embedding matrix, the first dimension $f$ was 631,261, which related to the total number of statements in our study. Hence function embedding matrix $\mathbf{F}$ would be shape of 631,261 $\times$ 150, where each row $\mathbf{F_{i}}$($i = 1, 2, 3, ... , f$) represented the code embedding for the $i$th function.

\vspace{0.1cm}\noindent\textbf{Statement Embedding Matrix} $\mathbf{S^{s \times d}}$: For statement-level code embedding matrix, same as contract-level and function-level, the first dimension $s$ corresponded to the total number of statements, which was 1,944,513 in our study. The shape of statement embedding matrix $\mathbf{S}$ would be 1,944,513 $\times$ 150, each row of the matrix represented the code embedding for a specific statement.

\subsection{Similarity Checking}
\label{sec:similarity}

We define the similarity checking methods in this step, which will be used in the following clone detection, clone-related bug detection, and contract validation tasks.

\vspace{0.1cm}\noindent\textbf{Definition} : Given two code fragments $C_1$ and $C_2$ , $e_1$ and $e_2$ are their corresponding code embeddings, we define the semantic distance as well as similarity between the two code snippets as below:

\begin{equation}
Distance(C_1, C_2)= \frac{Euclidean(e_1, e_2)}{\norm{e_1 }+ \norm{e_2}} \label{eq2}
\end{equation}
\begin{equation}
Similarity(C_1, C_2)= 1 - Distance(C_1, C_2) \label{eq3}
\end{equation}
Given any two code fragments $C_i$ and $C_j$, if their similarity score estimated above over a specific similarity threshold $\delta$, $C_i$ and $C_j$ are viewed as a clone pair. This similarity checking methods can be employed with vector space comparison and thus benefit ultimate tasks.


\subsection{Clone Detection, Bug Detection, and Contract Validation}
Based on the code embeddings we generated and the similarity checking methods we proposed, we are able to apply our approach to solve various tasks, i.e.,  clone detection, bug detection, and contract validation. For clone detection, we measure the similarity between two code fragments of smart contracts, and identify them as clone if the similarity score is above a pre-defined threshold. For bug detection, we search code fragments in our code base that are ``similar'' to the known bugs, then we identify the code snippets as buggy if its similarity score is over a pre-defined threshold. Moreover, for contract validation, when a developer complete a new smart contract, we also measure the similarity between it and the buggy statements we collected. If the similarity score is above a pre-defined threshold, the vulnerable statements can be identified in the new smart contract. Note that the threshold used for each of these three tasks can be different due to differences in the nature of these tasks. 


\section{Empirical Evaluation}
\label{sec:eval}
The main idea of our approach is based on code embedding and similarity checking for various similarity-based software engineering tasks. Herein, we evaluate how well our approach embeds code and checks similarity for the purposes of contract code clone detection, bug detection, and contract validation.

\subsection{Code Embedding Evaluation}
\begin{figure}
\centerline{\includegraphics[width=0.45\textwidth]{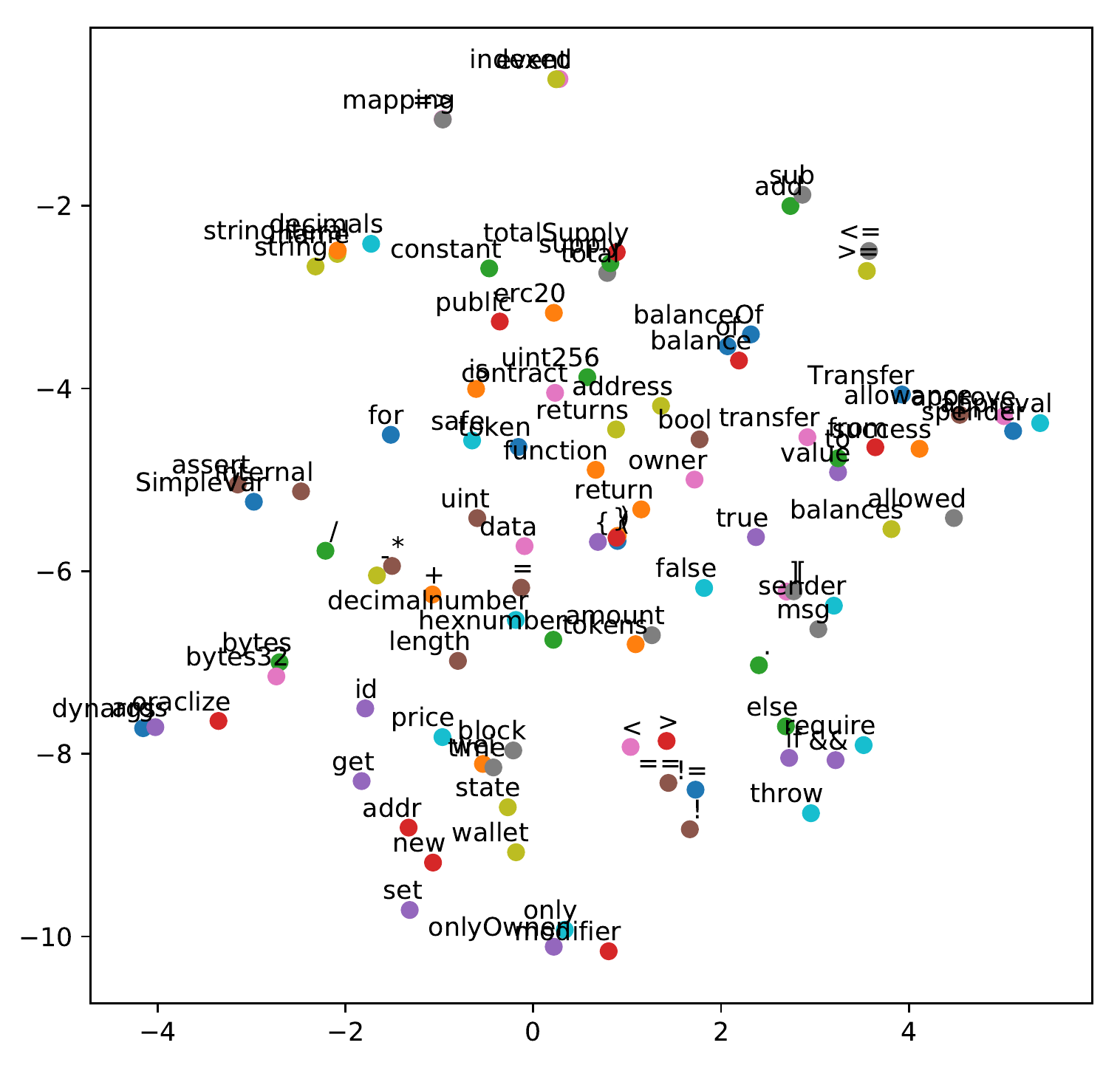}}
\vspace*{-5pt}
\caption{Result of Code Embeddings}
\label{fig:codeembeddings}
\end{figure}

As we have introduced in previous sections, representation learning maps a symbol to a real-valued, distributed vector.
the basic criterion of code embedding is that similar symbols should have similar representations. In particular, symbols that are similar in some aspects should have similar values in corresponding feature dimensions.
To demonstrate the effectiveness of our code embedding, we pick top 100 frequent tokens, then draw the embeddings for the tokens on a 2D plot using T-SNE algorithm, which are shown in Fig.~\ref{fig:codeembeddings}. Similar words that are close together in the vector space and are expected to be close in the 2D plot as well.

From the figure, we note that tokens sharing similar syntactic and lexical meaning are clustered together. For example, operators such as ``$+$'', ``$-$'', ``$*$'', ``$/$'', ``$>=$'', ``$<=$'' are grouped together, and tokens such as ``args'', ``dynargs'', and ``StringLiteral'', ``decimals'' are close to each other. This gives us confidence that high dimensional code representation can meaningfully capture co-occurrence statistics and distributed semantics for the tokens.

\subsection{Similarity Checking Evaluation}
To demonstrate the effectiveness of the similarity checking, we evaluate our approach with respect to three tasks: code clone detection, bug detection, and contract validation; and we compare the results with the following tools designed specifically for those tasks.

\begin{itemize}
\item Deckard \cite{jiang2007deckard}: a scalable, tree-based tool for source code clone detection. It has been widely used and extended to support the Solidity language, and we can compare with it on smart contract code clone detection.
\item SmartCheck \cite{tikhomirov2018smartcheck}: an extensive static analysis tool that can detect many kinds of vulnerabilities in smart contracts automatically.
It works on Solidity source code, and has been shown to outperform many other tools in terms of bugs detected. Hence in our study, we choose SmartCheck to compare the performance of our approach in detecting bugs and validating contracts.
\end{itemize}

In the following sections,  
we aim to answer the following \minor{six} key  research questions:
\begin{itemize}
    \item \textit{RQ-1:} How effective is our {\sc SmartEmbed} for detecting code clones within smart contracts?
    \item \textit{RQ-2:} How effective is {\sc SmartEmbed} for bug detection in smart contracts?
    \rev{
    \item \textit{RQ-3:} How effective is {\sc SmartEmbed} for distinguishing the bug fixes from the bugs?
    \item \textit{RQ-4:} How effective is the structural and semantic information added to {\sc SmartEmbed}?
    \item \textit{RQ-5:} How effective is {\sc SmartEmbed} for smart contract validation?
    \item \textit{RQ-6:} How efficient is {\sc SmartEmbed}?
    }
\end{itemize}


\subsection{\minor{RQ-1: Clone Detection Evaluation}}
\label{sec:cloneeval}

Code clones are common in software and can be considered useful or harmful depending on different circumstances.
They can appear more frequently in smart contracts than traditional software as smart contracts are irreversible and often intended to be self-contained, containing all the code implementing needed functionalities with little reference to other contracts. Maintaining smart contracts and managing duplications, redundancies, and inconsistencies are very important for contract quality assurance,
and the detection of contract code clones is an important first step. The nature of the task is similarity based and very suitable for our approach.

\subsubsection{Experimental Setup}
\label{sec:clonesetup}
Code clone detection is done through the vector space comparison via similarity checking, which is described in Section~\ref{sec:approach}. A similarity threshold governs whether two code fragments are viewed as clones. 
\rev{We evaluate the code clone detection at the contract level, function level as well as the statement level by using our approach}.

\begin{itemize}
\item Contract-level clone detection: As mentioned in Section~\ref{sec:approach}, each smart contract can be represented by a fixed dimensional vector. We construct a pairwise similarity matrix $M^{s \times s}$(in our case, M would be a 22718 $\times$ 22718 matrix, we removed 7 parsing error cases here), where each row and column corresponds to a smart contract,
and each cell $M_{ij}$ corresponds to the similarity score between smart contract $s_i$ and $s_j$. Given a similarity threshold $\delta$, if $M_{ij} > \delta(i \neq j)$, the corresponding smart contract $s_{i}$ and $s_{j}$ would be considered as a clone pair.
\item Function-level clone detection: Theoretically we could also construct a pairwise similarity matrix the same as the above, for all functions. However, due to the large number of functions, which was 631261, the complexity of computing the pairwise similarity between every pair of functions directly is too expensive. Hence in this evaluation, we randomly sample 200 smart contracts from our repository and use the functions in the 200 contracts, which contain 5307 functions in total, as clone queries. Following that, a pairwise similarity matrix $N^{s \times t}$ between the sampled 5307 functions and all of the functions in the whole contract set is generated (i.e., $N$ was a 5307 $\times$ 631261 matrix), where each cell $N_{ij}$ represented the similarity score between the sampled function $N_{i}$ and the function $N_{j}$. Same as the above, the associated functions $f_{i}$ and $f_{j}$ will be considered as a clone pair if $N_{ij} > \delta$.
\rev{
\item Statement-level clone detection: Same with function-level clone detection, since it is too expensive to calculate the pairwise similarity between every pair of statements directly, we extract all the statements within the aforementioned 200 sampled contracts, which contain 16,350 statements in total. Following that, we construct a pairwise similarity matrix $Q^{s \times t}$ between the sampled 16,350 statements and all of the statements in the whole contract set
(i.e., $N$ was a 16,350 $\times$ 1,944,513 matrix), where each cell $Q_{ij}$ represents the similarity score between the sampled statement $Q_{i}$ and the statement $Q_{j}$. Same as the above, the associated statements $s_{i}$ and $s_{j}$ will be considered as a clone pair if $Q_{ij} > \delta$.
}
\end{itemize}

\begin{table}[tbp]
    \caption{Code Clone Quantity Summary}
    \label{tab:clonequantity}
    \begin{center}
    \minor{
        \scriptsize{
            \begin{tabular}{|c|c|r|r|r|}
                \hline
                \textbf{Methods} & \textbf{Granularity} & \textbf{\# Cloned} & \textbf{\# Total} & \textbf{Clone} \\ & \textbf{level} & \textbf{lines} & \textbf{lines} & \textbf{ratio}  \\
                \hline
                \multirow{4}{*}{Deckard(1.0)} 
                & Original & 6623509 & 7329362 & 0.9039   \\
                & Contract & 4337582 & 7329362 & 0.5918   \\
                & Function & 24504 & 27045 & 0.9060 \\
                & Statement & 16448 & 18117 & 0.9079 \\
                \hline
                \multirow{3}{*}{SmartEmbed(1.0)} 
                & Contract & 2864673 & 7329362 & 0.3908 \\
                & Function & 23087 & 27045 & 0.8537 \\
                & Statement & 14774 & 18117 & 0.815 \\
                \hline
                \multirow{3}{*}{Deckard(0.95)} 
                & Original & 7054568 & 7329362 & 0.9625 \\
                & Contract & 5337860 & 7329362 & 0.7283 \\
                & Function & 26232 & 27045 & 0.9699 \\
                & Statement & 17548 & 18117 & 0.9685 \\
                \hline
                \multirow{3}{*}{SmartEmbed(0.95)} 
                & Contract & 6264136 & 7329362 & 0.8547 \\
                & Function & 24640 & 27045 & 0.9110 \\
                & Statement & 16760 & 18117 & 0.925 \\
                \hline
            \end{tabular}
        }
    }
    \end{center}
\end{table}

\begin{table*}
    \caption{Code Clone Quality and Overlapping Summary}
    \label{tab:clonequality}
    \begin{center}
    \rev{
        \begin{tabular}{|c|c|c|c|c|c|}
            \hline
            \textbf{Granularity} & \textbf{Similarity} & \textbf{Reported by} & \textbf{Reported by} & \textbf{Reported by} & \textbf{Overlap} \\
                        & \textbf{Threshold}  & \textbf{Deckard only} &  \textbf{both}      & \textbf{SmartEmbed only}   &  \textbf{ratio} \\
            \hline
            \multirow{2}{*}{Contract Level} & 1.0 & 1499308  & 2838274 & 26399 & 0.65 \\
            & 0.95 & 97140 & 5240720 & 1023416  & 0.82 \\
            \hline
            \multirow{2}{*}{Function Level} 
            & 1.0  & 1689   & 22815   & 272  & 0.92 \\
            & 0.95 & 1664   & 24568   & 72   & 0.93 \\
            \hline
            \multirow{2}{*}{Statement Level} 
            & 1.0  & 1933   & 14515   & 259  & 0.87 \\
            & 0.95 & 945    & 16603   & 157   & 0.93 \\
            \hline
        \end{tabular}
    }
    \end{center}
\end{table*}

\begin{figure}
\centerline{\includegraphics[width=0.22\textwidth]{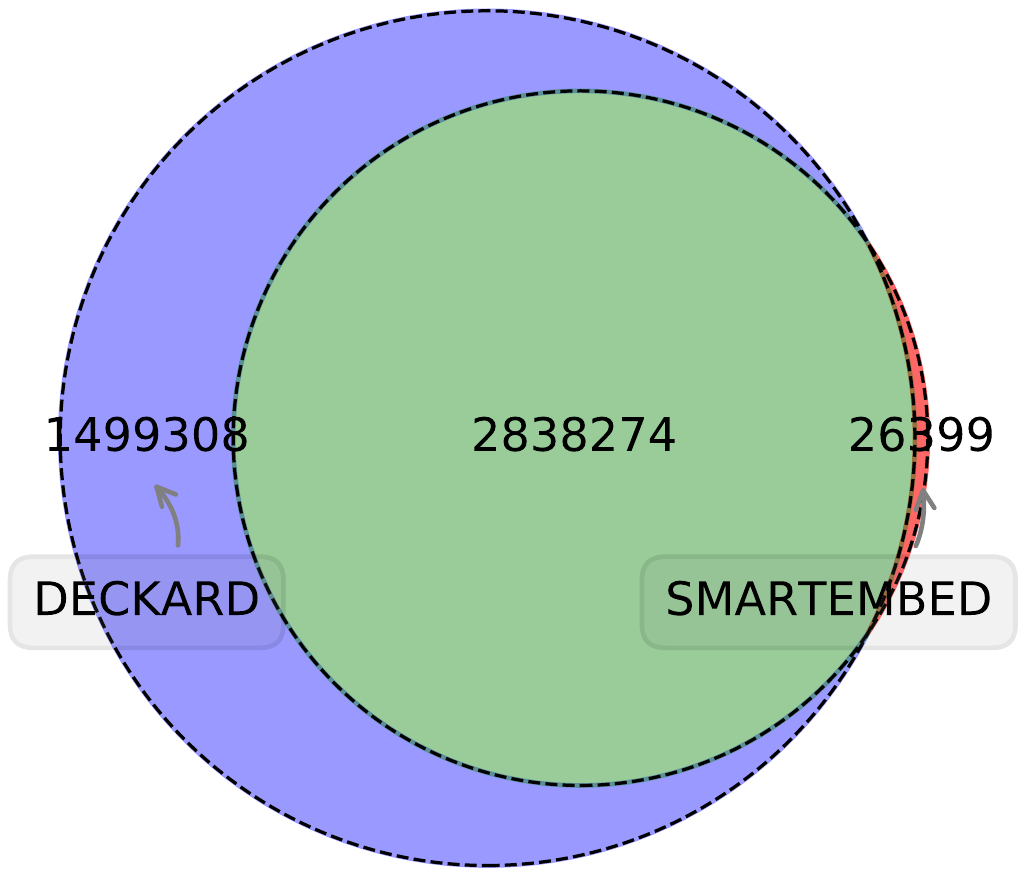}
			\includegraphics[width=0.23\textwidth]{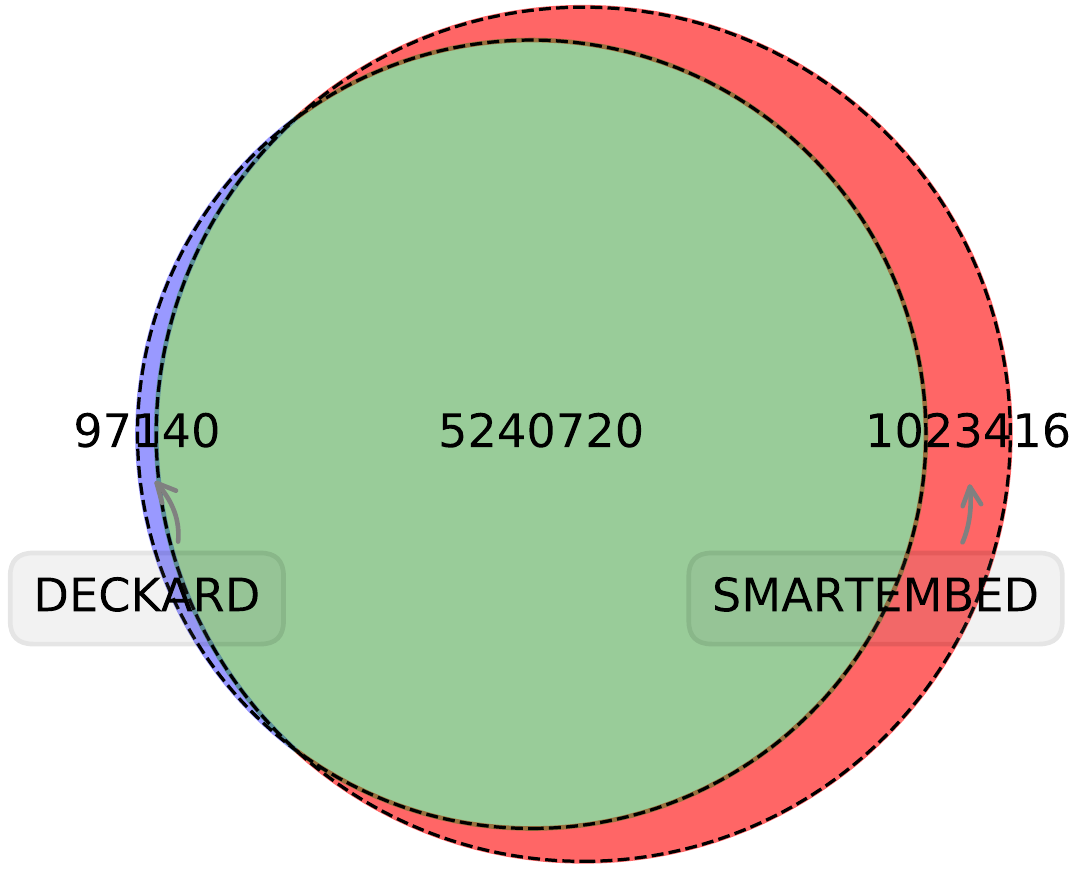}}
\caption{Venn Graph for Contract-Level Clones Detected by {\sc SmartEmbed} and Deckard with similarity threshold 1.0 (left) and 0.95 (right)}
\label{fig:clonesvenn-contract}
\end{figure}

\begin{figure}
\centerline{\includegraphics[width=0.247\textwidth]{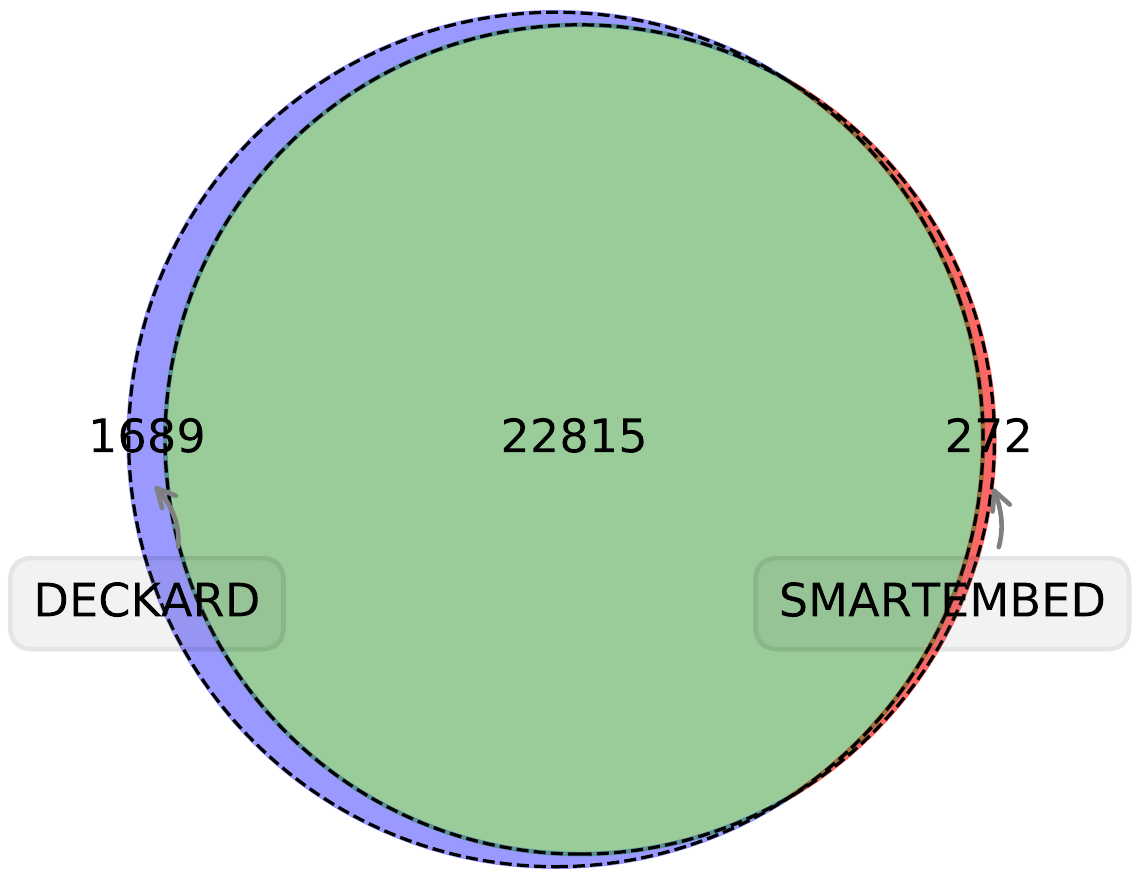}
			\includegraphics[width=0.245\textwidth]{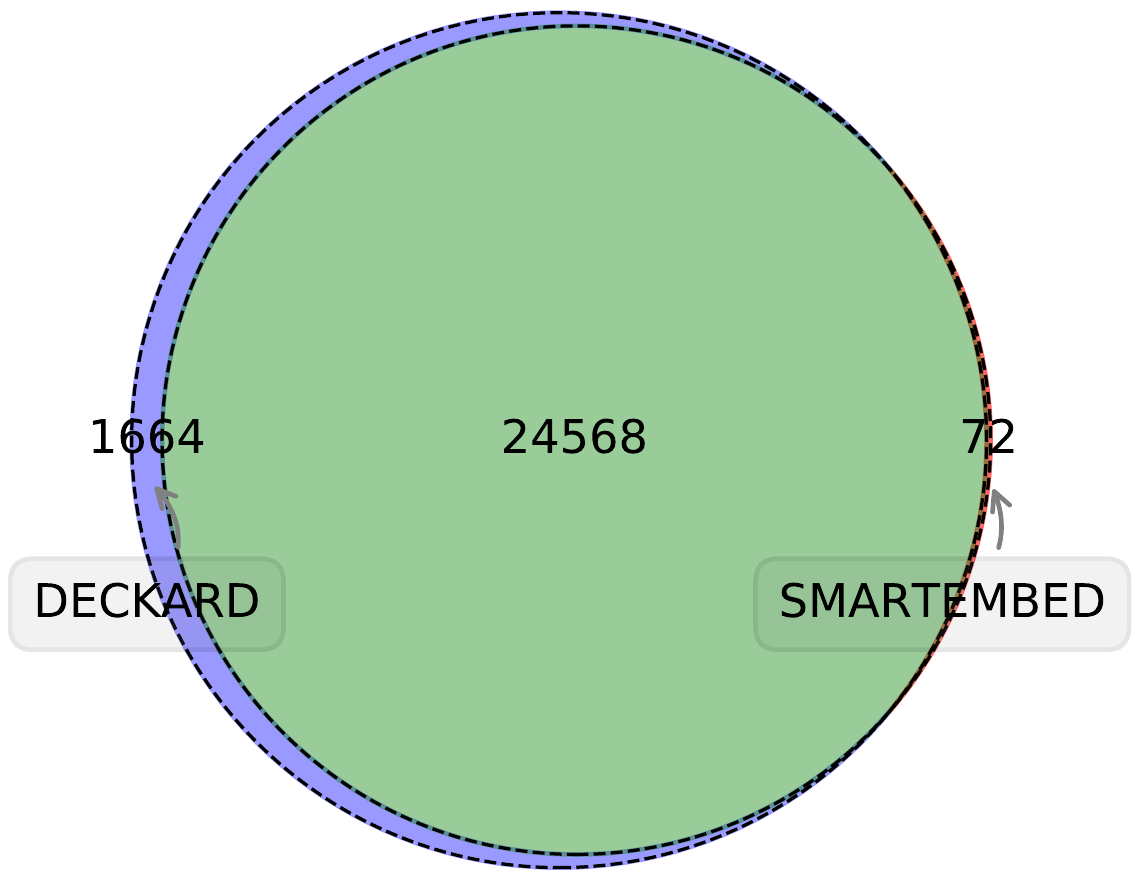}}
\caption{Venn Graph for Function-Level Clones Detected by {\sc SmartEmbed} and Deckard with similarity threshold 1.0 (left) and 0.95 (right) }
\label{fig:clonesvenn-function}
\end{figure}

\begin{figure}
\centerline{\includegraphics[width=0.247\textwidth]{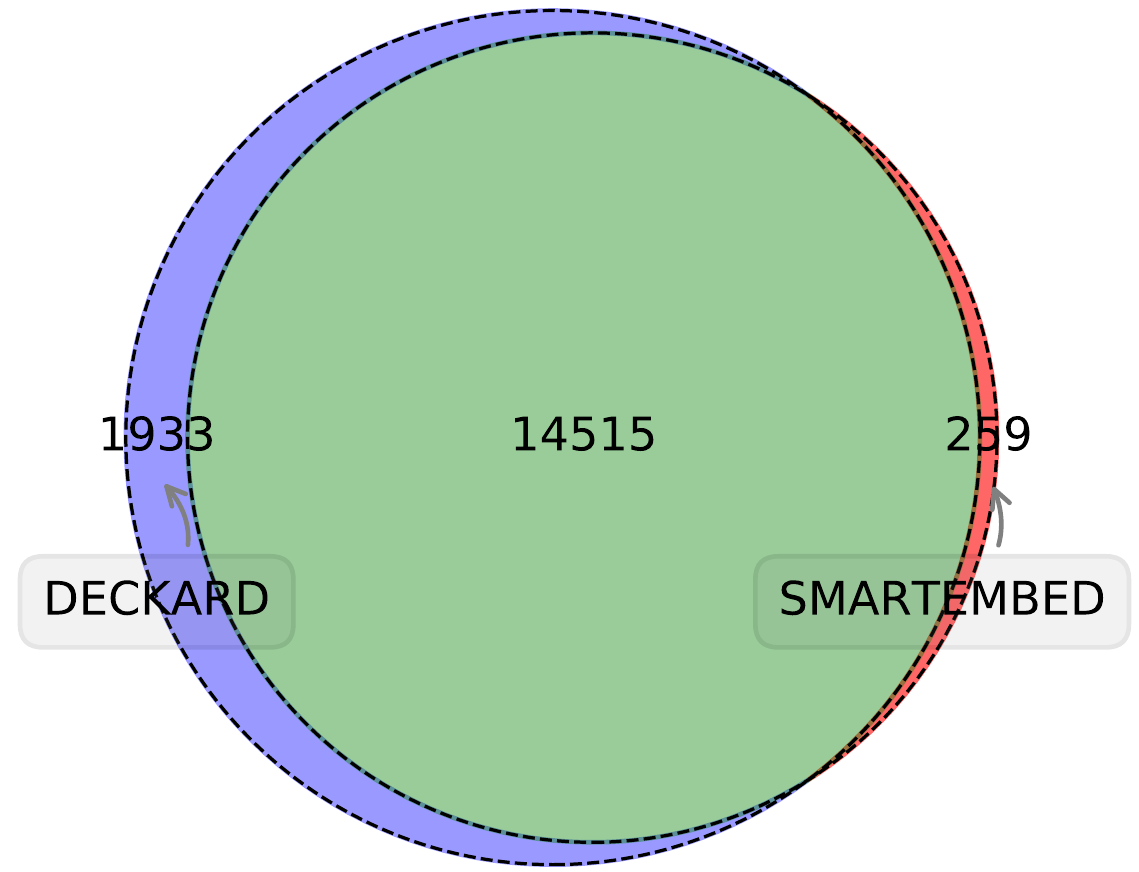}
			\includegraphics[width=0.245\textwidth]{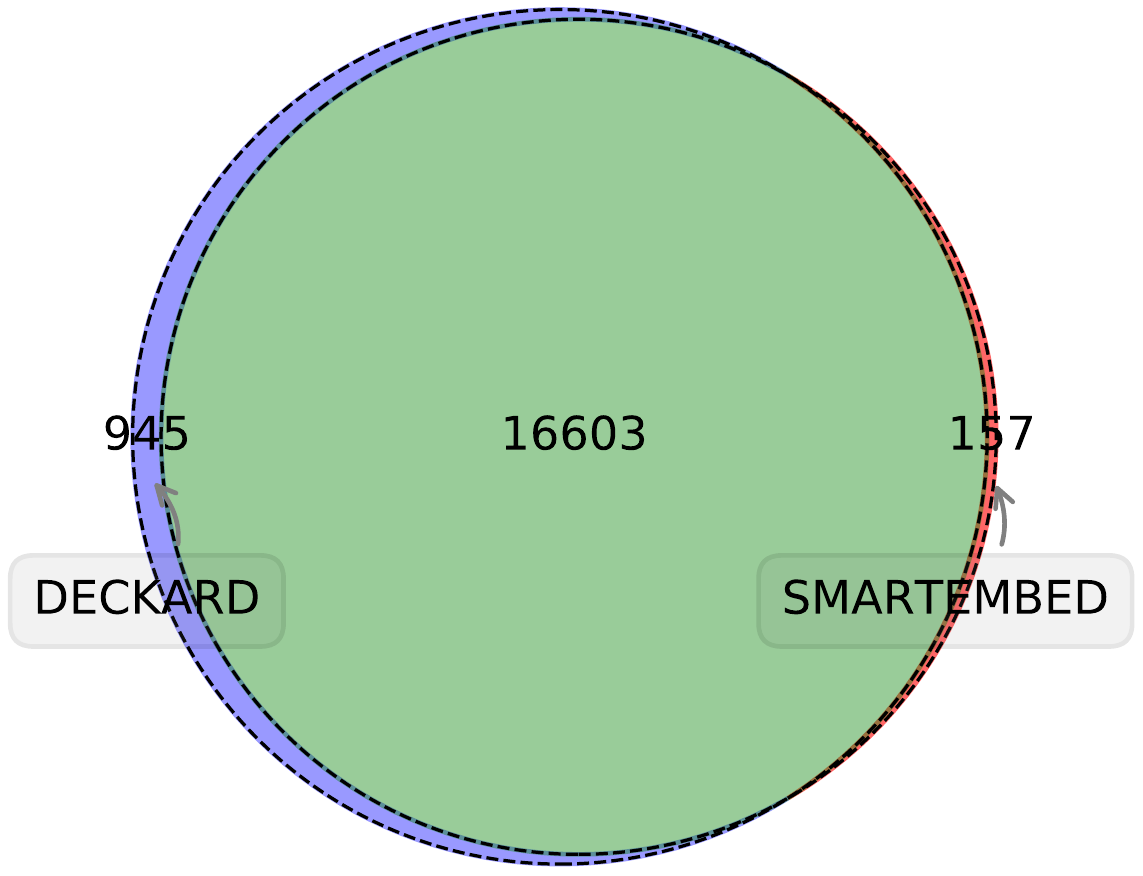}}
\caption{Venn Graph for Statement-Level Clones Detected by {\sc SmartEmbed} and Deckard with similarity threshold 1.0 (left) and 0.95 (right) }
\label{fig:clonesvenn-statement}
\end{figure}

\subsubsection{Experimental Results}
\rev{
To justify our approach on the task of code clone detection, we compare our results with those of Deckard (with its default settings) by the numbers of lines of code that are detected as clones. We set the similarity threshold to 1.0 and 0.95 for Deckard and {\sc SmartEmbed} respectively.\footnote{\rev{The definitions of similarity used in SmartEmbed and Deckard are not exactly the same: SmartEmbed is based on the embedding vectors (cf.~Section \ref{sec:similarity}); Deckard~\cite{jiang2007deckard} is based on tree structures. However, we simply assume the two are approximate of each other and treat them the same for easier comparison.}}
The results are summarized in Table~\ref{tab:clonequantity}. From the table, we can observe the following points.
}
\rev{
\begin{itemize}
    \item \textbf{There is a very high ratio of code clones among smart contracts.} By using Deckard with its default settings with similarity threshold 1.0, the code clones may involve more than 6.6 million lines of code, while the total lines in 22725 contracts are just 7.3 million, which means more than 90\% smart contracts on Ethereum are somehow cloned from others. The code clone ratio is even higher (more than 96\%) if we set the similarity threshold to 0.95. Since {\sc SmartEmbed} can detect code clones on contract-level, function-level and statement-level, we exclude the clone fragments in Deckard results that are smaller than a contract, function and statement respectively for a fair comparison. The clone ratios on both function-level and statement-level are consistent with the original clone ratio. We note that clone ratio drops at contract-level, this is because we just keep the results if the 
    whole contract is a clone, removing all the non-whole contract clones.  
    \item \textbf{{\sc SmartEmbed} report less clones overall than Deckard on different levels of granularity and similarity thresholds.}
    Regarding the {\sc SmartEmbed} results, the clone ratio was 0.39 and 0.85 at the contract-level with respect to similarity threshold 1.0 and 0.95 respectively. At the function-level, as mentioned in the previous subsection, we randomly sample 200 contracts which include 5,307 functions, involving 27,945 lines of code int total. {\sc SmartEmbed} detected 23,087 (85\%) and 24,640 (91\%) of them as clones with similarity threshold 1.0 and 0.95 respectively. Consistent with the function-level clone results, the clone ratio was 0.82 and 0.93 at statement-level with respect to the similarity threshold 1.0 and 0.95 respectively. 
    We argue that the main reason for this phenomenon is that \textbf{{\sc SmartEmbed} is more precise than Deckard in detecting clones}, this is because {\sc SmartEmbed} encodes both structural and some contextual semantic information, while Deckard only considers structural information. So, {\sc SmartEmbed} should have more constraints and detect less clones.
    \item \textbf{Most code clones detected by {\sc SmartEmbed} are also detected by Deckard.} To evaluate the quality of code clones reported by our approach, we count the numbers of lines of code in our results that overlap with clones reported by Deckard (assuming Deckard's results are accurate), the results are summariized in Table~\ref{tab:clonequality} (for both 1.0 and 0.95 similarity) and the Venn diagrams in Fig.~\ref{fig:clonesvenn-contract}, Fig.~\ref{fig:clonesvenn-function} and Fig.~\ref{fig:clonesvenn-statement} for the contract-level, function-level and statement-level respectively. We note that the overlap ratio is more stable at function-level, reflecting that {\sc SmartEmbed} is better in finding functional clones while tolerating non-essential syntactic differences. 
\end{itemize}
}
\rev{    
Regarding the relatively high clone ratio in smart contracts, we consider that the following reasons can be responsible for introducing clones:
\begin{itemize}
    \item One of the main reasons for introducing clones in smart contracts is the irreversibility of smart contracts stored in the Ethereum blockchain. Even when the same contract creator may want to evolve the contract code and create new versions of the smart contracts, the older versions are still kept visible in the blockchain. We consider such a scenario, and recount all the clones by creator addresses (i.e., if the detected clones are code belonging to a same creator, we do not report them), such clone results still report a considerable high clone ratio 51\% for similarity threshold 0.95 on contract level, reflecting the fact that cloning contracts across different creators is more common than usual software.
    \item ERC20 is the main technical standards for the implementation of tokens. The standardization allows contracts to operate on different tokens seamlessly, thus boosting interoperability between smart contracts. From the implementation perspective, ERC20 are interfaces defining a set of functions and events, such as \textit{totalSupply()}, \textit{balanceOf(address owner)}, \textit{transfer(address to, uint value)}. For every contract in our database, if the contract has implemented all the interfaces required by ERC20, it will be considered as an ERC20 contract. Finally, we find that 15,514 out of 22,725 (68.3\%) contracts contain the code blocks to support compliance to the ERC20 standard, reflecting that template contracts also plays an important role to cloning in Ethereum. 
\end{itemize}
}
\rev{The experimental results reveals homogeneous of the Ethereum ecosystem. Our clone detection results can benefit the smart contract community as well as individual Solidity developers in the following ways:
\begin{itemize}
    \item The relatively high ratio of code clones in smart contracts may cause severe threats, such as security attacks, resource wastage, etc. Finding such clones can enable significant applications such as vulnerability discovery (clone-related bugs) and deployment optimization (reduce contract size and duplication), hence contribute to the overall health of the Ethereum ecosystem. 
    \item Our work in identifying clones can also help Solidity developers to check for plagiarism in smart contracts, which may cause a huge financial loss to the original contract creator.
\end{itemize}
}

\begin{figure}[t]
    \centerline{\includegraphics[width=0.45\textwidth]{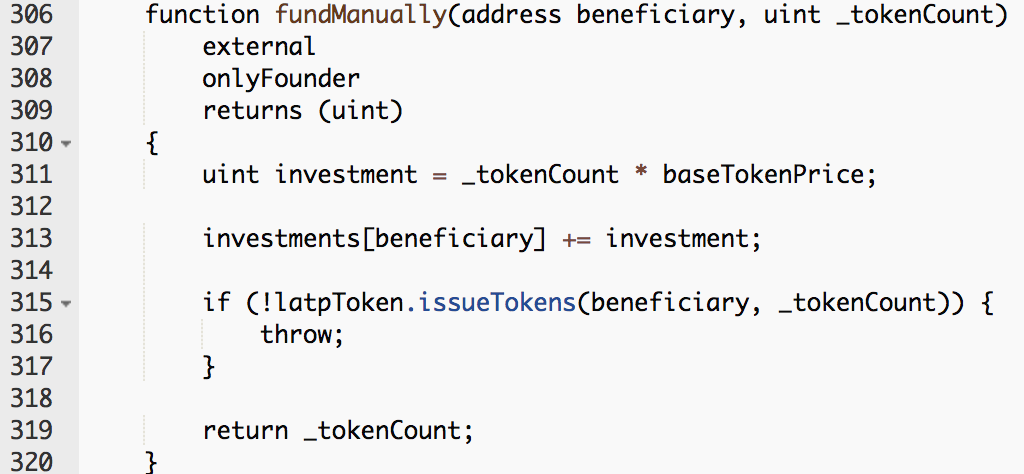}}
\caption{Example Pairs of SmartEmbed}
\label{fig:clonecode1}
\end{figure}

\begin{figure}[t]
    \centerline{\includegraphics[width=0.45\textwidth]{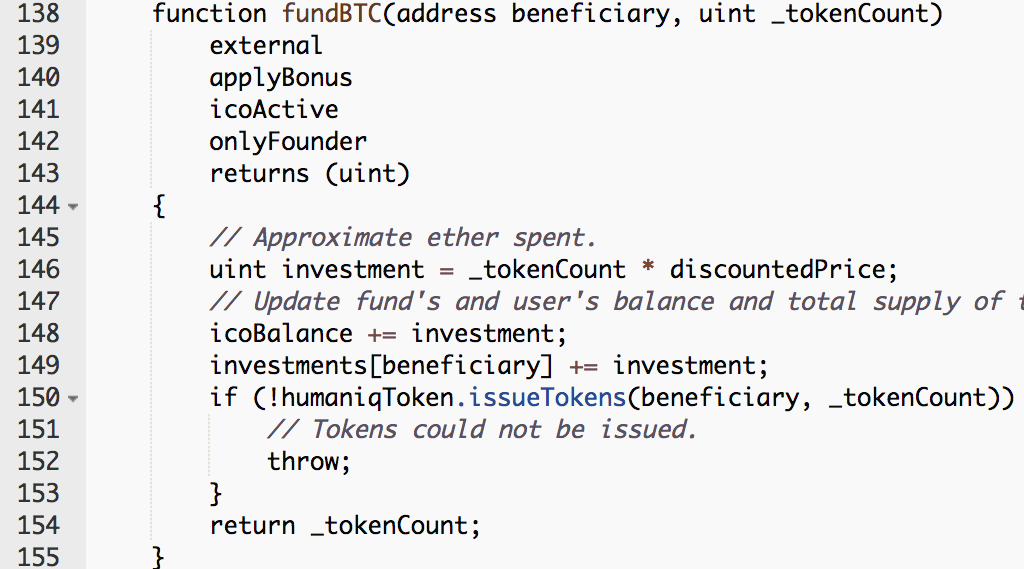}}
\caption{Example Pairs of SmartEmbed}
\label{fig:clonecode2}
\end{figure}

\begin{figure}[t]
    \centerline{\includegraphics[width=0.48\textwidth]{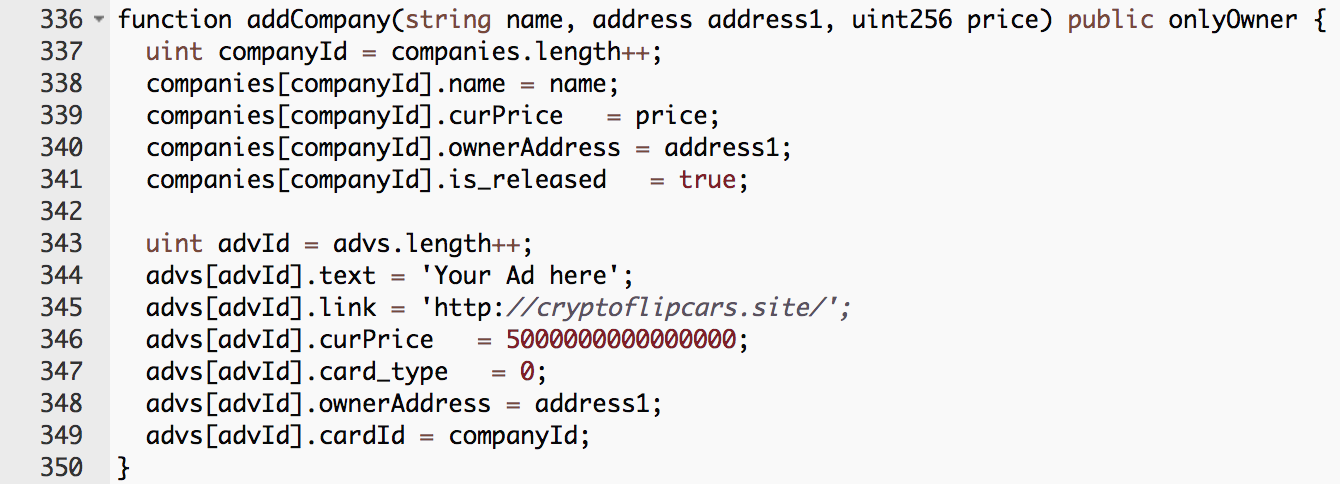}}
\caption{Example Pairs of Deckard}
\label{fig:notclonecode1}
\end{figure}

\begin{figure}[t]
    \centerline{\includegraphics[width=0.48\textwidth]{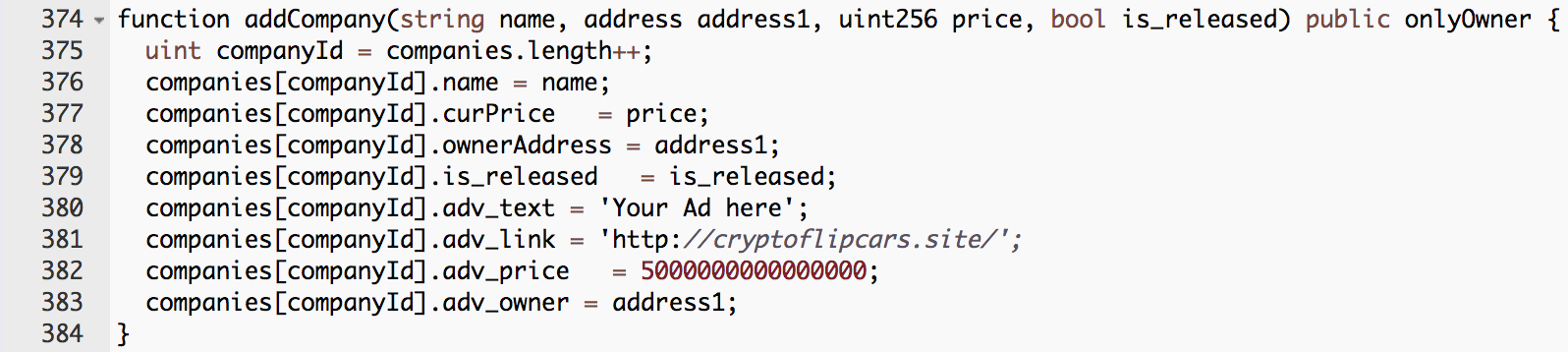}}
\caption{Example Pairs of Deckard}
\label{fig:notclonecode2}
\end{figure}

\subsubsection{Examples of clone detection}
\rev{
To compare the results of {\sc SmartEmbed} and Deckard, we have manually checked the clones detected by {\sc SmartEmbed} but not by Deckard. A sample code pair is shown in Fig.~\ref{fig:clonecode1} and Fig.~\ref{fig:clonecode2}.
The code pair has similar statements but some statements are added and modified, which can be considered as a type-III or even type-IV semantic clones \cite{roy2007survey} and are hard for Deckard to detect as it was designed for syntactic clones.
}

\rev{
We also manually checked the code clone pairs detected by Deckard but not by {\sc SmartEmbed}. A sample code pair is shown in Fig.~\ref{fig:notclonecode1} and Fig.~\ref{fig:notclonecode2}. Even though these two pieces of code are both functions about ``addCompany'', since they use different data structures, they are not considered as syntactic clones. This is because Deckard ignores the different identifier names in the code, which results in detecting this clone by accident.  
Regarding {\sc SmartEmbed}, it maintains these differences in identifier names, which increases the differences between associated code embedding vectors. This further justifies that {\sc SmartEmbed} is more precise in clone detection than Deckard.
}


\textbf{Answer to \textit{RQ-1:} How effective is our {\sc SmartEmbed} for detecting code clones within smart contracts?} - we conclude that {\sc SmartEmbed} is highly effective.

\subsection{\minor{RQ-2: Bug Detection Evaluation}}
\label{sec:bugeval}

\begin{table}
\caption{Vulnerable Smart Contracts}
\label{tab:buggycontracts}
\begin{center}
{\scriptsize
\begin{tabular}{|c|c|c|}
\hline
    {\bf Vulnerability} & {\bf Smart contract name} & {\bf Line num} \\
    \hline\hline
    \multirow{5}{*}{Overflow/Underflow}     & *SMT              & 206  \\ \cline{2-3}
                                  & *EthConnectPonzi  & 201  \\ \cline{2-3}
                                  & *BecToken         & 257  \\ \cline{2-3}
                                  & MESH              & 209  \\ \cline{2-3}
                                  & ethpyramid        & 217  \\ \hline
    \multirow{5}{*}{Blockhash/Timestamp}         & *SmartBillisons   & 554  \\ \cline{2-3}
                                  & *Ethraffle        & 94   \\ \cline{2-3}
                                  & *LuckyDoubler     & 118  \\ \cline{2-3}
                                  & KeberuntunganAcak & 124  \\ \cline{2-3}
                                  & Ethraffle\_v4b    & 92   \\ \hline
    \multirow{6}{*}{Implicit Visibility/HoneyPot}     & *Multiplicator    & 22   \\ \cline{2-3}
                                  & *PrivateBank      & 35   \\ \cline{2-3}
                                  & *KingOfTheHill    & 12   \\ \cline{2-3}
                                  & ETH\_VAULT        & 38   \\ \cline{2-3}
                                  & Simpson           & 25   \\ \cline{2-3}
                                  & RichestTakeAll    & 15   \\ \hline
    \multirow{6}{*}{Overpowered User/Owner CVE}   & *EthLendToken     & 236  \\ \cline{2-3}
                                  & *BitCoinRed       & 42   \\ \cline{2-3}
                                  & *Rubixi            & 18   \\ \cline{2-3}
                                  & NetkingToken      & 184  \\ \cline{2-3}
                                  & ZupplyToken       & 241  \\ \cline{2-3}
                                  & Toorr             & 42   \\ \hline\hline
    
    \multirow{2}{*}{\rev{Reentrancy}}   & *DAO              & 1013 \\ \cline{2-3}
                                  & MICRODAO          & 1001 \\ \hline
    \multirow{5}{*}{\rev{Gas Consumption/Gas Limit}}          & *Simoleon         & 61   \\ \cline{2-3}
                                  & *Penis             & 63   \\ \cline{2-3}
                                  & *FreeCoin          & 59   \\ \cline{2-3}
                                  & Polyion           & 102  \\ \cline{2-3}
                                  & Pandemica         & 50   \\ \hline
    \multirow{5}{*}{\rev{Incorrect Signature/Replay}}       & *MTC               & 211  \\ \cline{2-3}
                                  & *CNYToken          & 213  \\ \cline{2-3}
                                  & *GGoken            & 144  \\ \cline{2-3}
                                  & UGToken           & 140  \\ \cline{2-3}
                                  & CNYTokenPlus      & 180  \\ \hline
                               
    \multirow{6}{*}{\rev{TransferFlaw/ERC-20 Transfer}} &*UselessEthereumToken & 65 \\\cline{2-3}
                                    &*PhilcoinToken   &83    \\\cline{2-3}
                                    &*CosmosToken     &58    \\\cline{2-3}
                                    &*XmanToken       &61    \\\cline{2-3}
                                    &TacoToken       &120   \\\cline{2-3}
                                    &WinlastmileToken &104  \\\hline
                                    
    \multirow{6}{*}{\rev{Overflow/Batch Overflow}} & *TUPC     & 261 \\\cline{2-3}
                                &*WMCToken           & 193 \\\cline{2-3}
                                &*InsightChainToken  & 288 \\\cline{2-3}
                                &*NemoXXToken        & 259 \\\cline{2-3}
                                &FishOne            & 360 \\\cline{2-3}
                                &UpcToken           & 261 \\\hline
    
    \multirow{6}{*}{\rev{Unsafe Reverse/Verify Reverse}} & *CockMight & 61 \\\cline{2-3}
                            & *Collegecoin        & 53     \\\cline{2-3}
                            & *SynthornToken      & 58     \\\cline{2-3}
                            & *VilijavisShares    & 107    \\\cline{2-3}
                            & Frikandel          & 71     \\\cline{2-3}
                            & Virgo\_ZodiacToken   & 99     \\\hline  
                                          
\end{tabular}
}
\end{center}
\end{table}

To quickly duplicate some functionality, programmers usually copy and paste code, which can introduce clone-related bugs into programs.
It is also folklore that programmers often repeat similar bugs.
Such intuitions give the basis for similarity-based bug detection using our approach.
To pinpoint a bug accurately, we perform bug detection at the statement level of granularity. That is, for a given known buggy statement (simply called a bug), every statement in our code base whose similarity with respect to the bug exceeds a specific threshold is reported as a potential bug. As shown in the evaluation results later, compared with other analysis-based approach, our similarity-based approach can detect bugs similar to known ones across a large set of programs more efficiently and accurately, while analysis-based approach may detect more bugs in individual programs.

\subsubsection{Experimental Setup}
\rev{
To detect bugs, we need to collect some known buggy statements to construct the bug database. Although there are many contracts in the wild reported to be vulnerable (e.g., \cite{DAO2018,Parity2017}), there is a lack of a comprehensive list of references to pinpoint buggy statements in those contracts.
We collected a list of 52 known buggy smart contracts belonging to 10 kinds of common vulnerabilities. 
These vulnerabilities are from real world events (e.g., Reentracy, Honeypot, Replay, Gas Limit) \cite{DAO2018, Parity2017, TrailofBits2018}, previous research papers (e.g., Overflow/Underflow, Blockhash/Timestamp) \cite{luu2016making, atzei2017survey, li2017survey}
and/or the CVE reported by some organizations (e.g., Transfer Flaw, Batch Overflow, Verify Reverse) \cite{PeckShield-batch, PeckShield-transfer, PeckShield-allowAnyone}.
} 

\rev{
We then tried our best to pinpoint buggy statements in those contracts by inspecting research papers, web articles, and community discussions. A list of vulnerable smart contracts and their vulnerabilities are summarized in Table~\ref{tab:buggycontracts}. For each vulnerable smart contract in the table, one or more associated buggy lines are identified.
We divide the 52 vulnerable smart contracts into two groups: 32 smart contracts marked with * are used for the bug detection evaluation, the other 20 are saved for the contract validation evaluation later. For the bug detection evaluation, 63 buggy statements are collected from the 32 vulnerable contracts. We create our bug database from the 63 buggy statements by using code embedding described in Section~\ref{sec:approach}. That is, for each buggy statement, we compose a numerical vector by summing up the vectors for all relevant tokens in the statement. Each statement is thus mapped to a vector of 150 dimensions.
Since we have 63 buggy statements, a bug embedding matrix $\mathbf{V^{63 \times 150}}$ is constructed and serves as our bug database.
}

\begin{figure}
\centerline{\includegraphics[width=0.4\textwidth]{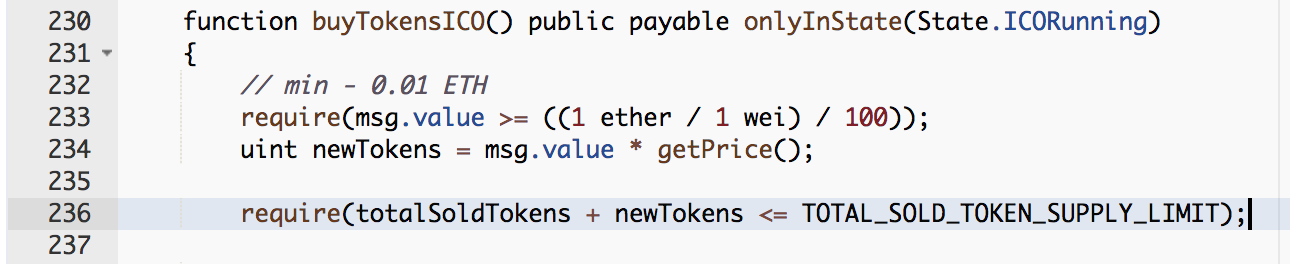}}
\caption{Real bug:EthLendToken@236}
\label{fig:type2-real}
\end{figure}

\begin{figure}
\centerline{\includegraphics[width=0.4\textwidth]{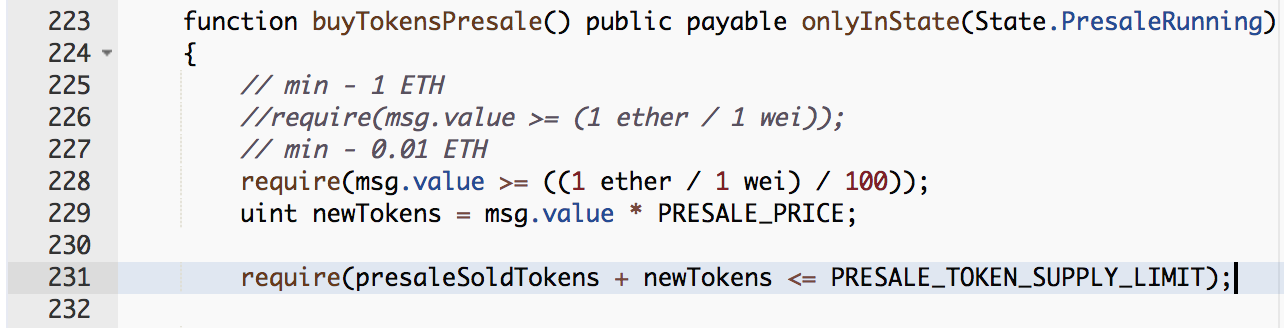}}
\caption{Candidate bug:UHubToken@231}
\label{fig:type2-candidate}
\end{figure}

\begin{figure}
\centerline{\includegraphics[width=0.4\textwidth]{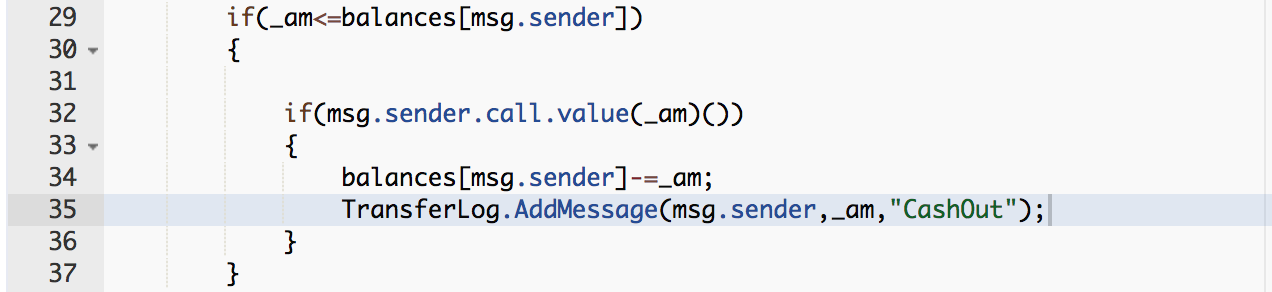}}
\caption{Real bug:PrivateBank@29-37}
\label{fig:type3-real}
\end{figure}

\begin{figure}
\centerline{\includegraphics[width=0.4\textwidth]{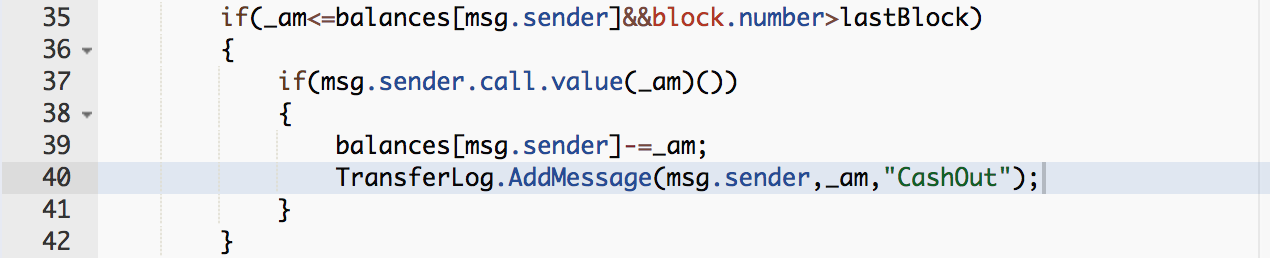}}
\caption{Candidate bug:ETH\_FUND@35-42}
\label{fig:type3-candidate}
\end{figure}

The setting for bug detection herein is that, for each buggy statement embedding $\mathbf{V_{i} \in V}$ in our bug database (simply called a bug), we need to identify every possible statement $\mathbf{S_{j} \in S}$ that is in the set of all statements in the contracts we collect from the Ethereum blockchain and similar to the given bug.
Given a similarity threshold $\delta$, if the similarity score estimated between $\mathbf{S_{j}}$ and $\mathbf{V_{i}}$ is over $\delta$, then $\mathbf{S_{j}}$ will be reported as a potential bug similar to $\mathbf{V_{i}}$.
We perform such bug detection to report bug candidates for every bug in our bug database.
Following that, we validate each candidate bug to see whether it involves an actual bug or not by manually checking.
To be more specific, we compare bug candidate lines reported by our approach with the real bug lines, the candidate bugs will be validated if one of the following conditions was satisfied:

\begin{itemize}
    \item The bug statements contain the exact identical code fragments same as the real bugs, which can be considered as type-I clone-related bugs.
    
    \item The bug candidates involve syntactically equivalent fragments as real bugs, with some variations in identifiers, literals or types, which can be viewed as type-II clone-related bugs. A sample pair is shown in Fig.~\ref{fig:type2-real} and Fig.~\ref{fig:type2-candidate}.
    \item The candidate bug lines involve syntactically similar code with inserted, deleted or updated statements, which can be considered as type-III or type-IV clone-related bugs. A sample pair is shown in Fig.~\ref{fig:type3-real} and Fig.~\ref{fig:type3-candidate}.
\end{itemize}

If the bug candidate is an actual clone-related bug, then it is counted as validated in Table~\ref{tab:bugtypeprecision} and Table~\ref{tab:bugprecision}. To demonstrate the advantages of {\sc SmartEmbed} in clone-related bug detection, we also compare it with the detection results of SmartCheck.

\begin{table}[t]
\caption{\rev{Bug Detection Precision Summary for Various Clone Types for Similarity Threshold 0.90}}
\label{tab:bugtypeprecision}
\begin{center}
\rev{\scriptsize
\begin{tabular}{|c|c|c|c|c|}
    \hline
    {\bf Clone Type} & {\bf \# bugs reported} & {\bf \# bugs validated} & {\bf precision} & {\bf ratio} \\
    \hline\hline
    Type-I  & 116 & 116   & 100\% & 8.8\%\\
    \hline
    Type-II & 989 & 989   & 100\% & 75.4\%\\
    \hline
    Type-III/IV & 69 & 58   & 84.1\% & 5.3\%\\
    \hline
    Not-Clones & 137 & 0   &  0\% & 10.5\%\\
    \hline
    Total & 1,311 & 1,163   &  88.7\% & 100\%\\
    \hline
\end{tabular}
}
\end{center}
\end{table}

\subsubsection{Experimental Results}
\rev{
For different types of clones, the bug detection results of {\sc SmartEmbed} are summarized in Table~\ref{tab:bugtypeprecision}. By setting the similarity threshold to 0.90, we count the number of reported bugs as well as validated bugs with respect to each clone type (i.e., type-I, type-II, type-III/type-IV). If the bug candidate does not belong to any of these clone types, it is identified as Not-Clones. From the table, we can observe the following points.
\begin{itemize}
    \item Most of the bug candidates reported by {\sc SmartEmbed} are Type-II clones. This reflects that solidity developers do introduce the clone-related bugs by copying and pasting source code from somewhere else.
    \item {\sc SmartEmbed} can achieve 100\% precision for detecting Type-I and Type-II clone-related bugs. This is because Type-I and Type-II clones do not involve structural changes and can be easily identified.
    \item The performance of {\sc SmartEmbed} drops for detecting the Type-III/IV clones. To identify the Type-III/IV clone-related bugs, we need to decrease the similarity threshold, which may also introduce more false positive cases at the same time. 
\end{itemize}
}

\rev{
The bug detection results of {\sc SmartEmbed} with respect to different similarity threshold are summarized in Table~\ref{tab:bugprecision}.}
For each specific similarity threshold $\delta$ in the table, we show the number of reported bug candidates (i.e., the number of statements in our set of contracts that have a similarity higher than $\delta$ to some bug in our bug database), and the number of bugs validated by manual checking together with the precision.
From Table~\ref{tab:bugprecision}, we can see that:
\begin{itemize}
\item The precision of  {\sc SmartEmbed} increases as the similarity threshold increases. For thresholds higher than 0.96,  {\sc SmartEmbed} can have a 100\% precision.
\rev{
\item The lower the $\delta$ is, the more statements may be reported as potential bugs. When the similarity threshold is set to 0.91, {\sc SmartEmbed} reports 1,052 statements as potential bugs, while maintaining a high precision of 95\%.
\item When the similarity threshold is set to 0.90, {\sc SmartEmbed} reports 1,311 potential bugs, 1,163 of them are validated as real bugs. The precision of {\sc SmartEmbed} drops to 88.7\%. This is reasonable because smaller similarity threshold will bring in more noises and hence incur more challenges for detecting clone related bugs. It also signals that setting the similarity threshold between 0.90 and 0.91 may be a good choice for the bug detection task.
}

\end{itemize}
\rev{
Since it is too expensive to run SmartCheck on all the 20k+ contracts, we only run it on the manually validated contracts associated with the 1,163 statements. SmartCheck automatically checks a given contract for predefined vulnerability patterns and highlights the lines of code containing the vulnerabilities. For a fair comparison, we limit SmartCheck to the bug patterns we collected in Table~\ref{tab:buggycontracts}. \minor{SmartCheck only reported 697 out of 1163 statements as bugs,} which shows the advantage of our approach in detecting clone-related bugs.
}


\begin{table}[t]
\caption{\rev{Bug Detection Precision Summary for Various Clone Similarity Thresholds}}
\label{tab:bugprecision}
\begin{center}
\rev{
\begin{tabular}{|c|c|c|c|}
    \hline
    {\bf Threshold} & {\bf \# bugs reported} & {\bf \# bugs validated} & {\bf precision} \\
    \hline\hline
    1.0  & 116 & 116   & 100\% \\
    \hline
    0.99 & 156 & 156   & 100\% \\
    \hline
    0.98 & 248 & 248   & 100\% \\
    \hline
    0.97 & 322 & 322   & 100\% \\
    \hline
    0.96 & 437 & 437   & 100\% \\
    \hline
    0.95 & 582 & 572   & 98.3\% \\
    \hline
    0.94 & 736 & 723   & 98.2\% \\
    \hline
    0.93 & 875 & 858   & 98.0\% \\
    \hline
    0.92 & 1014 & 983  & 96.9\% \\
    \hline
    0.91 & 1107 & 1052  & 95.0\% \\
    \hline
    0.90 & 1311 & 1163  & 88.7\% \\
    \hline
\end{tabular}
}
\end{center}
\end{table}


\subsubsection{Examples of bug detection}

We manually checked some bugs reported by {\sc SmartEmbed} but not by SmartCheck.
Some types of bugs, such as ``Honeypots'' in Table~\ref{tab:buggycontracts} can not be effectively checked by SmartCheck.

For example, the function {\tt multiplicate()} above is the only function that does allow a call from anyone other than the owner. It looks like by sending a value  higher than the current balance of the contract it is possible to withdraw the full balance from the contract. Both statements in line 7 and 9 try to reinforce the idea that {\tt this.balance} is somehow credited after the function is finished. However, this is a trap since the {\tt this.balance} is automatically updated before the {\tt multiplicate()} function is called. So {\tt if(msg.value>=this.balance)} is never true unless {\tt this.balance} is initially zero.

\begin{lstlisting}[language=Java, caption=MultiplicatorX3 example]
contract MultiplicatorX3 {
    ...
    function multiplicate(address adr)
    public
    payable
    {
        if(msg.value>=this.balance)
        {
            adr.transfer(this.balance+msg.value);
        }
    }
}
\end{lstlisting}

Encoding such a bug type into tools like SmartCheck would require extra efforts in defining the bug specification, while our approach can just take the sample bug and automatically generate embeddings to recognize similar bugs.
Of course, this advantage of our approach relies on good embeddding of all relevant structural and semantic information of code, which will be a continuing research direction in the future.

\textbf{Answer to \textit{RQ-2:} How effective is {\sc SmartEmbed} for bug detection in smart contracts?} - we conclude that {\sc SmartEmbed} is very effective for clone-related bug detection in a large set of smart contracts.

\subsection{\minor{RQ-3: Practical Analysis}}
\rev{
Considering the cloning rate in Ethereum is remarkably higher than the traditional software, a key problem with code cloning is that the original piece of code should ideally be fixed in every copy of its later versions. \minor{Herein we perform a practical analysis to verify whether {\sc SmartEmbed} can distinguish bug fixes from the original buggy statement.}
}

\subsubsection{Experimental Setup}
\rev{
Because the code file of deployed contracts is immutable, hence when a bug is identified in a smart contract, the developer should deploy a fixed version to the Ethereum blockchain. 
For each buggy smart contract in our bug database, we manually investigated the contract creation history of the contract creator to see if there is a fixed version contract for the specific buggy statement. 
Finally we found that 5 out of 52 buggy smart contracts include a fixed version. We pinpointed the fixed statement and estimated the similarity score  between the buggy statement and its corresponding fixed statement. 
}
\subsubsection{Experimental Results}
\rev{
\minor{The practical analysis results of {\sc SmartEmbed} are summarized in Table~\ref{tab:effective_analysis}}. 
A similarity score is calculated between the buggy statement and its corresponding fixed statement. From the table, we can see that:
\begin{itemize}
    \item By setting the similarity threshold to 0.90, all the fixed smart contracts can be correctly identified by {\sc SmartEmbed} as not vulnerable. Even though the original version and fixed version are very similar, {\sc SmartEmbed} can effectively identify the real clone-related bugs and neglect those fixed ones. This is because {\sc SmartEmbed} focuses on statement-level for bug detection, any small fixes within the buggy statement will result in different code embedding vectors, which will also reduce the similarity scores. 
    \item There is a significant drop of similarity scores between the fixed version contracts and the original ones. This further justifies the ability of {\sc SmartEmbed} to separate the real buggy statement and fixed statement. 
\end{itemize}
}

\begin{table}[t]
\caption{\rev{\minor{Practical Analysis}}}
\label{tab:effective_analysis}
\begin{center}
\rev{
\begin{tabular}{|c|c|c|c|}
    \hline
    {\bf Contract Name} & {\bf Similarity (fixed)} & {\bf Report Bug (0.90)}  \\
    \hline\hline
    BitcoinRed & 0.798 & False \\
    \hline
    CockMight & 0.883 & False \\
    \hline
    FishOne & 0.733 & False \\
    \hline
    WMCToken & 0.726 & False \\
    \hline
    XmanToken & 0.668 & False \\
    \hline
\end{tabular}
}
\end{center}
\end{table}

\subsubsection{\minor{Bug and Bug Fix Examples for Practical Analysis}}
\begin{figure}[t]
    \centerline{\includegraphics[width=0.52\textwidth]{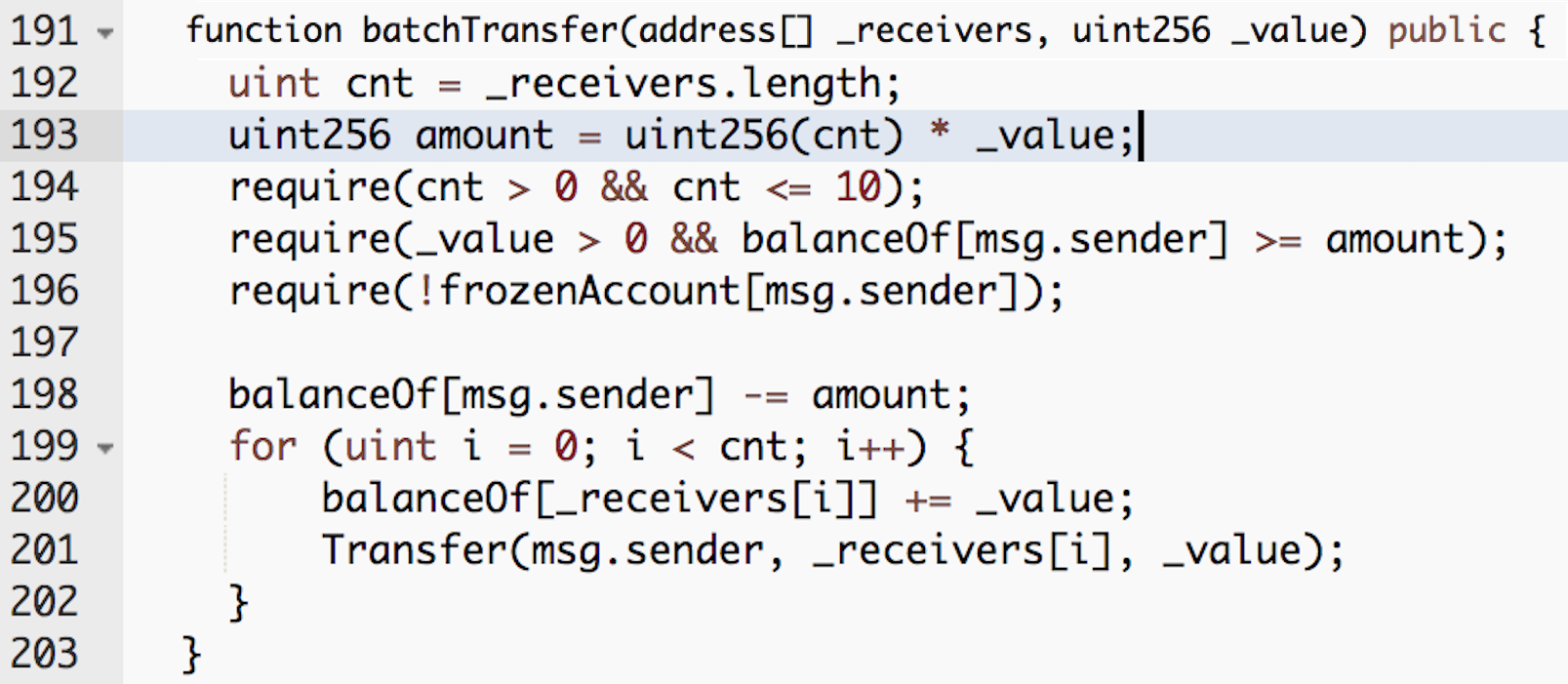}}
\caption{Original Version of WMCToken@193}
\label{fig:original_version}
\end{figure}

\begin{figure}[t]
    \centerline{\includegraphics[width=0.52\textwidth]{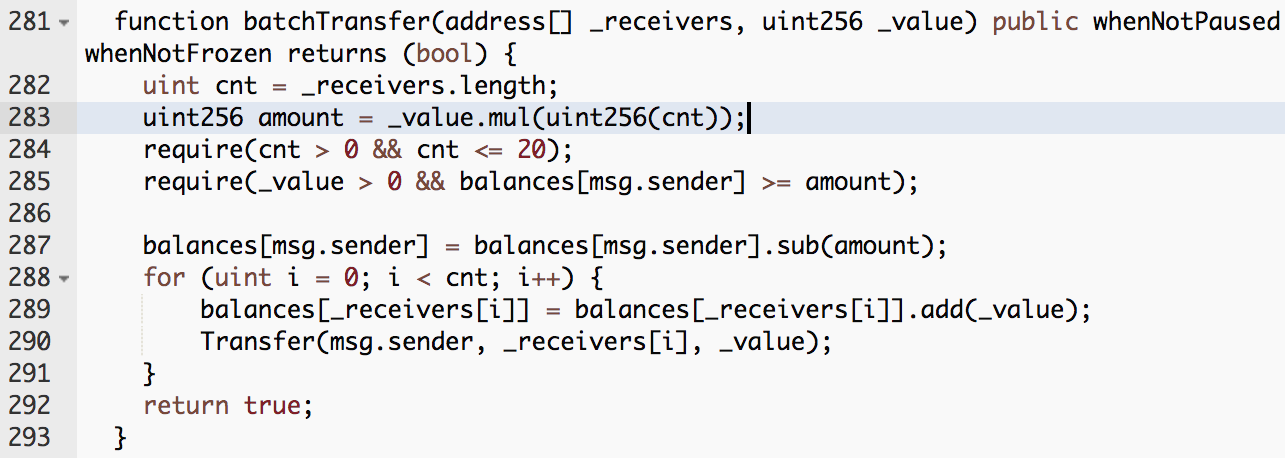}}
\caption{Fixed Version of WMCToken@283}
\label{fig:fixed_version}
\end{figure}

\rev{
We show a pair of original buggy statement and its corresponding fixed statement in Fig.~\ref{fig:original_version} and Fig.~\ref{fig:fixed_version}. 
As illustrated in Fig.~\ref{fig:original_version}, the function \textit{batchTransfer()} makes multiple transactions simultaneously. By passing several transferring addresses and amounts by the caller, the function would conduct some checks then transfer tokens by modifying balances. However, overflow might occur in line 193, \textit{uint256 amount = uint256(cnt) * \_value}, if \textit{\_value} is a huge number. It will make \textit{amount} become a small value rather than \textit{cnt} times of \textit{\_value}, then transfers out tokens exceeding \textit{balances[msg.sender]}.
For the fixed version of \textit{batchTransfer()} function in Fig.~\ref{fig:fixed_version}, the buggy statement is updated to \textit{uint256 amount = \_value. mul(uint256(cnt))}, herein, the contract creator compute the multiplication by using secure mathematical operations such \textit{SafeMath}. 
The change in the buggy statement as well as the function signatures reduce the similarity score between the buggy statement and the fixed statement.
}

\rev{
\textbf{Answer to \textit{RQ-3:} How effective is {\sc SmartEmbed} for distinguishing the bug fixes from the bugs?} - we conclude that {\sc SmartEmbed} is very effective for distinguishing the bug fixes from the clone-related bugs.
}

\subsection{\minor{RQ-4: Ablation Analysis}}
\label{sec:ablation}

\begin{table*}[t]
	\caption{\rev{Ablation Analysis}}
	\label{tab:ablation}
	\begin{center}
		\begin{tabular}{|c||c|c|c||c|c|c|}
			\hline
			\multirow{2}{*}{threshold} & \multicolumn{3}{|c||}{\bf SmartEmbed} & \multicolumn{3}{|c|}{\rev{\bf BasicEmbed}} \\\cline{2-7}
			& {\# bugs reported} & {\# bugs validated} & {precision}    & {\# bugs reported} & {\# bugs validated (sampled)} & {precision}  
			
			\\\hline\hline
			1.0  & 116 & 116   & 100\%   & 32,264 & 13 / 246   & 5.3\% \\
			\hline
			0.99 & 156 & 156   & 100\%   & 32,264 & 13 / 246   & 5.3\%  \\
			\hline
			0.98 & 248 & 248   & 100\%   & 32,265 & 13 / 246   & 5.3\% \\
			\hline
			0.97 & 322 & 322   & 100\%   & 32,268 & 13 / 246   & 5.3\%  \\
			\hline
			0.96 & 437 & 437   & 100\%   & 32,296 & 13 / 246   & 5.3\% \\
			\hline
			0.95 & 582 & 572   & 98.3\%   & 32,322 & 13 / 246  & 5.3\% \\
			\hline
			0.94 & 736 & 723   & 98.2\%   & 32,408 & 13 / 247  & 5.3\% \\
			\hline
			0.93 & 875 & 858   & 98.0\%  & 33,708 & 13 / 259   & 5.0\%  \\
			\hline
			0.92 & 1014 & 983  & 96.9\%  & 34,073 & 13 / 263   & 4.9\%  \\
			\hline
			0.91 & 1107 & 1052  & 95.0\% & 37,061 & 14 / 291   & 4.8\%  \\
			\hline
			0.90 & 1311 & 1163  & 88.7\% & 37,601 & 15 / 300   & 5\%  \\
			\hline
		\end{tabular}
	\end{center}
\end{table*}

\rev{
When we perform the bug detection, one main novelty of {\sc{SmartEmbed}} is adding details of structural (containment and neighbouring) and semantic (data-flow) information based on our serialization of parse trees. 
For example, we added the chain of ancestors in ASTs to capture sequence derivations and function signatures to capture the diverse neighbourhood relations of nodes.
As shown in Section~\ref{sec:bugeval}, this tree-based embedding technique is quite accurate and effective for bug detection in a large set of smart contracts. To verify the effectiveness of the structural and semantic information added to {\sc SmartEmbed}, we perform an ablation analysis with respect to the bug detection task.
}

\subsubsection{Experimental Setup}
\rev{
For the ablation analysis, we compare {\sc SmartEmbed} with one of its incomplete variants, named {\sc BasicEmbed}. 
Different from {\sc SmartEmbed}, {\sc BasicEmbed} removes all the structural and semantic relations from the statement tokenization results, and only keeps the simple statement token sequence. 
By going through the same steps of normalization, code embedding learning and embedding matrix building process, we can construct a new code embedding model for {\sc BasicEmbed}. 
Following that, for each bug statement in Table~\ref{tab:buggycontracts}, we apply {\sc BasicEmbed} to the bug detection task via similarity checking. 
}
\subsubsection{Experimental Results}
\rev{
The bug detection results of {\sc BasicEmbed} and {\sc SmartEmbed} are summarized in Table~\ref{tab:ablation}. 
Due to the very large number of bugs reported by {\sc BasicEmbed}, which is more than 30k+, manually validating all these potential bugs is too expensive. Herein this evaluation, we randomly sampled 300 contracts and validated these contracts manually. From the table, we have the following observations.
\begin{itemize}
    \item The total number of bugs reported by {\sc BasicEmbed} is very large, which is over 30k. At the same time, the overall precision of {\sc BasicEmbed} is only around 5\%, which means the majority of the bugs reported by {\sc BasicEmbed} are false positives. This also reflects that by simply extracting the token sequence of the statement is not accurate enough for the bug detection task.
    \item Regarding the precision of different similarity thresholds, {\sc SmartEmbed} stably and substantially outperforms {\sc BasicEmbed}, which reflects that the structural and semantic information have a major influence on the overall performance. This verifies the effectiveness and necessity of adding structural and context information based on parse trees.
    \item 87\% of the bugs reported by {\sc BasicEmbed} have a similarity threshold of 1.0, which means most of the bugs reported by {\sc BasicEmbed} are type-I clone-related bugs.
    This is because without considering the context of the statement, code clones with respect to a single buggy statement can be easily identified in other smart contracts.
    It further supports our claims that the structural and semantic relations convey much valuable information.
\end{itemize}
}
\subsubsection{Bug Detection Example for the Ablation Analysis}

\rev{
We manually checked some buggy statements that have a large number of clones reported by {\sc BasicEmbed}. For example, {\sc BasicEmbed} reported 10,679 potential bugs with respect to the following buggy smart contract.
}
\begin{lstlisting}[language=Java, caption=Rubixi example]
contract Rubixi {
    ...
    address private owner;
    function DynamicPyramid() { 
        owner = msg.sender; 
    }
    function collectAllFees() { 
        owner.send(collectedFees); 
    }
    ...
}
\end{lstlisting}
\rev{
The function above name {\tt DynamicPyramid} should be {\tt Rubixi}. The wrong name gives permissions to anyone to invoke the {\tt DynamicPyramid} function to become the owner of the contract and withdraw fees from it. If the function had the same name as the contract Rubixi, then the Ethereum virtual machine would automatically block access from anyone except the contract creator.
This bug happened at some point of time during the development of the contract: the contract name was changed from DynamicPyramid into Rubixi, but the programmers forgot to change the name of the constructor accordingly.
}

\rev{
The buggy statement of this smart contract is pinpointed at line 5, which is \textit{owner = msg.sender}. However, without considering context information, this simple statement can be easily identified in many other smart contracts with the exact identical code tokens, and most of these reported bugs are false positive cases. This is the reason for the extremely large number of bugs and very low precision by using {\sc BasicEmbed}. For using {\sc SmartEmbed}, we can encode the context of a statement, such as the function signatures \textit{function DynamicPyramid} and contract ancestor node \textit{Rubixi} into the code embedding vector, which can effectively reduce the false positive rate and identify the real bugs in other smart contracts.
}

\rev{
\textbf{Answer to \textit{RQ-4:} How effective is the structural and semantic information added to {\sc SmartEmbed}?} - we conclude that the structural and semantic information added to {\sc SmartEmbed} do have significant benefits for its overall performance.
}

\subsection{\minor{RQ-5: Contract Validation Evaluation}}
\label{sec:valeval}

\begin{table*}
	\caption{Confusion Matrix Summary}
	\label{tab:confusionmatrix}
	\vspace*{-12pt}
\rev{
	\begin{center}
		\begin{tabular}{|c|c|c|}
			\hline
			SE(0.95)/SE(0.90)/SE(0.85)/SC  & True Bugs & True Non-Bugs   \\ \hline
			Predicted Bugs     & 27 / 36 / 45 / 25  & 0 / 8 / 116 / 278          \\ \hline
			Predicted Non-Bugs & 18 / 9 / 0 / 20  & 2812 / 2804 / 2696 / 2534                 \\ \hline
		\end{tabular}
	\end{center}
}
\end{table*}

\begin{table}[tbp]
	\caption{Contract Validation Summary}
	\label{tab:contractvalidation}
	\rev{
	\begin{center}
		\begin{tabular}{|c|c|c|c|c|}
			\hline
			& SE(0.95) & SE(0.90) & SE(0.85) &  SC   \\ \hline
			Precision      & 100\%        & 81.8\%     & 28.1\%    & 8.3\%       \\ \hline
			Recall         & 60\%       & 80.0\%     & 100\%       & 55.6\%      \\ \hline
			F1             & 75\%       & 80.9\%     & 43.7\%      & 14.4\%      \\ \hline
			FPR            & 0.0\%      & 0.3\%      & 4.1\%       & 9.9\%       \\ \hline
			FNR            & 40\%       & 20\%       & 0\%         & 44.4\%      \\ \hline
		\end{tabular}
	\end{center}
	}
\end{table}

Because a smart contract is immutable once it is deployed onto the blockchain,
it would be better to ensure its correctness in its pre-deployment phase.
The objective of the experiment here is to test the capability of {\sc SmartEmbed} in catching all bugs in a smart contract that are similar to known bugs,
so as to help validate the correctness of the contract.
Although not a formal verification tool, our approach can grow its capability in validating a smart contract, as it is easily extensible to incorporate new known bugs into our bug database to check whether a smart contract contains similar bugs.

\subsubsection{Experimental Setup}
\label{sec:validationsetup}
To help validate a given contract, for each statement $s$ in the contract, we generate a 150 dimensional vector for $s$ based on our model and query it against all the bugs in our bug database $\mathbf{V^{63 \times 150}}$. If the similarity between $s$ and any bug in our bug database exceeds a threshold $\delta$ ($\delta$ is set to 0.95, 0.90 \& 0.85 for this task), $s$ can be reported as a potential bug.

\rev{
To assess the effectiveness of our approach, we took the 20 smart contracts without * in Table~\ref{tab:buggycontracts} for test.
Also, a list of ``bug-free'' smart contracts can help to assess false positive and false negative rates. Therefore, we collected 20 audited smart contracts from Zeppelin, one of the most popular security audit firms. Each vulnerability discovered on them is automatically considered as a false positive.
There are a total of 2857 statements associated with these 40 smart contracts (20 buggy and 20 bug-free); 45 statements from the 20 buggy contracts are labelled as bugs.
We performed bug detection on these smart contracts by using both our {\sc SmartEmbed} approach (SE) and SmartCheck (SC).
The confusion matrix with respect to the bug reports generated by SE with three different similarity thresholds (0.95, 0.90 and 0.85) and SC are summarized in Table~\ref{tab:confusionmatrix}. We also calculated the Precision, Recall, F1 score, FPR (false positive rate), and FNR (false negative rate) based on the confusion matrix and show the metrics in Table~\ref{tab:contractvalidation}.
}


\subsubsection{Experimental Results}
From Table~\ref{tab:confusionmatrix} and Table~\ref{tab:contractvalidation}, it can be seen that:
\begin{itemize}
    \item The majority of the bugs can be checked with our approach, and our approach can identify clone-related bugs more accurately than SmartCheck, which is consistent with bug detection evaluation results.
    \rev{
    \item By using our approach with the similarity threshold 0.90, the number of false positives was 8 and it decreased to 0 with the similarity threshold 0.95. SmartCheck reported far more false positives than ours. Since SmartCheck can check more kinds of bug patterns, it is worth noting that, for a fairer comparison, we only enabled the bug types listed in Table~\ref{tab:buggycontracts} for SmartCheck. When other types of vulnerabilities were disabled, SmartCheck still had a 9.9\% false positive rate; its FPR would be overwhelmingly higher if all bug types were enabled.
    }
    \rev{
    \item The number of clone-related bugs discovered by our approach increased from 27 to 36 with decreasing similarity thresholds from 0.95 to 0.90.
    A potential explanation is related to a common practice by developers who may do code cloning but make changes to the clones for various reasons.
    Such a practice may cause some cloned code to become dissimilar to each other, which would need lower thresholds to detect them.
    \item  The false negatives decreased to 0 when we set the similarity threshold to 0.85, which means all the bugs can be identified by our approach using this threshold. At the same time, the false positives reported by our approach increased to 116, but still far less than the results generated by SmartCheck. Looking at the F1 score of this similarity threshold, our approach is still much better than SmartCheck.
    }
\end{itemize}

\rev{
\textbf{Answer to \textit{RQ-5:} How effective is {\sc SmartEmbed} for smart contract validation?} - our results show that {\sc SmartEmbed} is effective in capturing bugs similar to known ones with low false positive rates. Our future work will also continue to enrich the bug database with more real bugs and improve the embeddings.
}



\subsection{\minor{RQ-6: Time Cost Analysis}}
\label{sec:timecost}


The time cost of {\sc SmartEmbed} is mostly for the training of code embeddings and the vector similarity checking, and is dependent on the sizes of contract codebase and bug database.
To analyze the complexity of our proposed approach, we need to measure the time complexity in the computation of similarity as defined in Eqn.(2)(3).
For our machine containing an Intel Xeon CPU E5-2640 v4 @ 2.40GHz, the training of code embedding took about a day for our dataset.
The average time for a pairwise similarity calculation between two code snippets, as defined in Equation (2) and (3) (Sec.~\ref{sec:similarity}) is around 250ns.
\rev{
We estimated the time by applying Deckard, {\sc SmartEmbed} and SmartCheck \minor{service tool} for clone detection, bug detection and contract validation tasks respectively. \minor{We use the same server described above for testing,} it took on average 79.2ms and 416.3ms to check a single smart contract by using Deckard and SmartCheck respectively.
Regarding {\sc SmartEmbed}, for clone detection, computing the pairwise similarity matrix $M$ ($M$ was a 22718$\times$22718 matrix) took on average 6.05s, checking each smart contract only cost 0.26ms. For bug detection, all statements in our contract codebase are queried against our bug embedding matrix, computing the similarity matrix $N$ ($N$ was a 1944513$\times$63 matrix) took on average 53.22s, checking each smart contract cost 2.3ms. For contract validation, a given contract is queried against our bug embedding matrix, which took on average 4.7ms.
}

\rev{
\textbf{Answer to \textit{RQ-6:} How efficient is {\sc SmartEmbed}?} - The query for a clone or a bug using {\sc SmartEmbed} is efficient for practical uses.
}

\section{DISCUSSION}
\label{sec:disc}

We selected several smart contract projects from Github, then contacted the Solidity developers by sending clone reports and bug reports generated by {\sc SmartEmbed} for these projects. For clone detection, we reported the most similar smart contracts' url on Etherscan associated with its similarity score. For bug detection, we reported the exact bug line and associated bug type. Some developers expressed interest in using our tool.

\begin{enumerate}[(1)]
\item \vspace{0.1cm}\noindent {\bf Clone Detection} - Compared to  Etherscan's ``find similar contract'' function, which can only find ``Exact Match'' contracts, our tool is more flexible which can report code clone on contract level, function level or even statement level governed by a similarity threshold. One practitioner responded, ``\textit{If the tool works with individual functions then that might be useful. I would give you a shout out on Twitter}''. Another developer commented, ``\textit{The clone detection isn't useful to me, but I could believe it would be useful to authors of widely cloned contracts, such as cryptokitties or FOMO3D}.''

\item \vspace{0.1cm}\noindent {\bf Bug Detection} - With the help of our techniques, developers could quickly check for vulnerabilities and improve confidence in the reliability of a contract. ``\textit{It is nice to have such a tool to identify vulnerable bugs in smart contract, I probably will give it a try}''. However, there are also some developers who mentioned that the bug report is not useful, ``\textit{one intractable problem I found was that in smart contracts, everything is dangerous, and you can't judge whether a contract is secure without understanding intent - any insecure pattern can be correct in the context of a contract designed to do that. }''

\end{enumerate}
\rev{
According to developers' comments, we have implemented {\sc SmartEmbed}\footnote{http://www.smartembed.net} as a standalone web application tool \cite{gao2019smartembed}.
Solidity developers can copy and paste their contract source code to the web application to find repetitive contract code and clone-related bugs in the given contract.
The source code of SmartEmbed and contract data used in our experiments can be found in our Github repository\footnote{https://github.com/beyondacm/SmartEmbed}. 
}

\rev{
Some developers also suggested publishing the tool as an extension and enhancement to Etherscan so that developers who have already been familiar with Etherscan can easily utilize the tool, which can facilitate broader adoption of the tool and easier collections of new bugs.
\minor{Since a lot of Solidity developers use the web IDE Remix to develop, deploy, and test a smart contract, developers also suggested integrating the tool as a plugin into an IDE (e.g., \minor{Remix and Visual Studio Code})} 
to help detect clones and bugs early in development. 
The efficiency of SmartEmbed's similarity checking step (excluding the embedding steps), as shown in Section~\ref{sec:timecost}, can be sufficient in supporting the uses in IDE in real-time when developers are writing their code.
We will follow such suggestions to improve the tool in the near future.
}



\section{Threats to Validity}
\label{sec:threats}

\vspace{0.1cm}\noindent {\bf Internal Validity.} Code representations used for code embedding have significant effects on the embedding outcome and the downstream applications. 
\rev{The ways we calculated the code embedding for each code snippet is intuitive, which may bias our approach for detecting clones of different code sizes.
There are a lof of related work have explored different ways to represent code and embed more semantic information into the code vectors,
} 
such as paths in control flow graphs\cite{defreez2018path}, paths in ASTs\cite{alon2018general}, dynamic execution traces\cite{wang2017dynamic}, API sequences and usage contexts \cite{nguyen2017exploring}, and many others. We will try to employ different code embedding techniques for the same tasks in the future.


\rev{
\vspace{0.1cm}\noindent {\bf Data Validity.} We collected 22,725 solidity smart contracts with source code through Etherscan for our experiment. It is not complete as the number of smart contracts on Ethereum grows faster recently and the number of contracts on Etherscan is almost doubled, over 40,000 already, not to mention many other contracts that do not provide source code. In the future, we can retrain our model and gain a better code representation model with the enlarged Solidity source code data set, and may even extend the embedding techniques to Solidity bytecode.
In addition, due to the lack of a comprehensive list of Ethereum contract vulnerabilities, the number of buggy contracts we collected is relatively small. Our bug database currently contains 52 buggy contracts covering 10 different bug types that are more relevant for Solidity smart contracts, ignoring bug types that may be common for other programming languages. The selected contracts may not be sufficiently diverse or representative of all contracts, and there can be a lot of false negatives if applying our approach to detect bug types not included in our bug database.
We will keep expanding both our code base and bug database in the near future.
}

\vspace{0.1cm}\noindent {\bf External Validity.}  We validated the clone-related bugs detected by {\sc SmartEmbed} only from the SmartCheck benchmark. One of the threat is that SmartCheck can also have the false negative as well as false positive cases, hence the results may be biased and incomprehensive.  There currently exists other security analysis tools to find bugs in smart contract, such as Oyente \cite{luu2016making}, Mythril \cite{Mueller2018}, Gasper \cite{chen2017under} and Securify \cite{tsankov2018securify}. We plan to do more large-scale evaluations with these tools in the near future. We also acknowledge that the sample size of the user study is not sufficient, we plan to get more feedback about our tool from practitioners in the future.


\section{Summary}
\label{sec:con}
We have proposed a new approach, {\sc SmartEmbed}, based on structural code embedding and similarity checking for clone detection, bug detection and contract validation tasks on smart contracts.
We have evaluated our approach with more than 22,000 Solidity smart contracts from the Ethereum blockchain. For clone detection, {\sc SmartEmbed} can effectively identify many instances of repetitive solidity code where the clone ratio is around 90\%, and more semantic clones can be detected accurately by our tool than Deckard. For bug detection, {\sc SmartEmbed} can identify more than 1000 clone-related bugs based on our bug databases efficiently and accurately, which can enable efficient checking of smart contracts with changing code and bug patterns. Such capabilities of {\sc SmartEmbed} can be useful for facilitating contract validation in practice.


\balance
\bibliographystyle{unsrt}
\bibliography{samples}

\end{document}